\documentclass[preprint,12pt, sort&compress,a4paper]{elsarticle}

\usepackage{ifthen}
\newboolean{draft}
\newboolean{arXiv}
\newboolean{external} \setboolean{external}{true}

\setboolean{draft}{false}
\setboolean{arXiv}{true}

\ifthenelse{\boolean{draft}}{\usepackage{todonotes}
}{\usepackage[disable]{todonotes}
}

\presetkeys{todonotes}{inline}{}

\ifthenelse{\boolean{arXiv}}{}{\usepackage[mathlines]{lineno}
\linenumbers
}

\usepackage[T1]{fontenc}
\usepackage[vmargin={20mm,25mm},hmargin={25mm,20mm}]{geometry}
\usepackage{graphicx}
\usepackage{subcaption}
\usepackage{caption}
\usepackage{amsmath}
\usepackage{xspace}
\usepackage{xparse}
\usepackage{xstring}
\usepackage{booktabs}
\usepackage{siunitx}
\usepackage{tikz}
\usetikzlibrary{calc,patterns,angles,quotes,backgrounds}

\definecolor{AzureishWhite}{HTML}{dee7f4} \definecolor{Eggshell}{HTML}{F4EBDE} \definecolor{ChineseWhite}{HTML}{DEF4E0} \definecolor{PinkLace}{HTML}{F4DEF2}

\definecolor{maxivorange}{HTML}{ff8b00}
\definecolor{maxivgreen}{HTML}{126018}
\definecolor{essblue}{HTML}{8faadc}
\definecolor{myyellow}{HTML}{ffc000}

\makeatletter
\if@twocolumn
\newcommand{\whencolumns}[2]{
#2
}
\else
\newcommand{\whencolumns}[2]{
#1
}
\fi
\makeatother

\newcommand{\figwidth}{\whencolumns{0.45}{1.0}\linewidth}

\renewcommand\eqref[1]{Eq.~(\ref{#1})\xspace}

\newcommand\tabref[1]{Table~\ref{#1}\xspace}
\newcommand\figref[1]{Fig.~\ref{#1}\xspace}
\newcommand\Figref[1]{Fig.~\ref{#1}\xspace}
\newcommand\secref[1]{Sect.~\ref{#1}\xspace}

\newcommand{\TwoLambda}{\ensuremath{2\Lambda_{th}}\xspace}
\newcommand{\TenTenTen}{\ensuremath{(\num{10} \times \num{10} \times \num{10})\,\unit{\cm\cubed}}\xspace}

\newcommand{\TenByTen}{\numproduct{10 x 10}}
\newcommand{\TenByTenCm}{\TenByTen\,\unit{\cm\squared}\xspace}
\newcommand{\ZeroOneByTenCm}{\ensuremath{(\num[round-precision=1]{0.1} \times \num {10})\,\unit{\cm\squared}}\xspace}
\newcommand{\ZeroZeroOneByTenCm}{\ensuremath{(\num[round-precision=2]{0.01} \times \num {10})\,\unit{\cm\squared}}\xspace}

\AtBeginDocument{\def\thefnote{\myfnsymbol{fnote}}}
\makeatletter
\def\myfnsymbol#1{\expandafter\@myfnsymbol\csname c@#1\endcsname}
\def\@myfnsymbol#1{\ifcase #1\or $\#$\or $\#\#$\else \@ctrerr\fi}
\def\fntext[#1]#2{\g@addto@macro\@fnotes{\refstepcounter{fnote}\elsLabel{#1}\def\thefootnote{\thefnote}\global\setcounter{footnote}{\c@fnote}\footnotetext{#2}}}
\makeatother

\newif\ifapp
\appfalse

\ifthenelse{\boolean{external}}{\usetikzlibrary{external}
\tikzexternalize
\NewCommandCopy{\oldtodo}{\todo}
\RenewDocumentCommand\todo{om}{\tikzexternaldisable \IfNoValueTF{#1}{\oldtodo{#2}}{\oldtodo[#1]{#2}}
\tikzexternalenable }
}{}

\usepackage[pdftex,
colorlinks,
linkcolor={red},
citecolor={essblue},
urlcolor={essblue}
pdfkeywords={moderators, neutronics}]{hyperref}

\title{A concept of a para-hydrogen-based cold neutron source for simultaneous high flux and high brightness}

\author[JCNS]{Alexander Ioffe\corref{cor1}}

\cortext[cor1]{Corresponding author.\\ \hspace*{4ex}E-mail address: \href{mailto:a.ioffe@fz-juelich.de}{a.ioffe@fz-juelich.de}~(A.\,Ioffe)}

\affiliation[JCNS]{organization={Juelich Centre for Neutron Science~JCNS at Heinz Maier-Leibnitz Zentrum~(MLZ)},
addressline={Lichtenbergstr.\,1},
postcode={85748},
city={Garching},
country={Germany}}

\author[JCNS]{Petr Konik\fnref{PK}}

\fntext[PK]{Current address: Neutron Spectroscopy Laboratory,
HUN-REN Centre for Energy Research, Konkoly-Thege Mikl{\'o}s {\'u}t,
29--33, 1121, Budapest, Hungary}

\author[MAXIV]{Konstantin Batkov}

\affiliation[MAXIV]{organization={MAX\,IV Laboratory, Lund University},
addressline={Fotongatan 2},
postcode={22484},
city={Lund},
country={Sweden}}

\begin{document}

\begin{frontmatter}

\begin{abstract}

A novel concept of cold neutron source employing chessboard or
staircase assemblies of high-aspect ratio rectangular para-hydrogen
moderators with well-developed and practically fully illuminated
surfaces of the individual moderators is proposed.

An analytic approach for calculating the brightness of para-hydrogen
moderators is introduced. Because brightness gain originates from a
near-the-surface effect resulting from the prevailing single collision
process during thermal-to-cold neutron conversion, high-aspect ratio
rectangular cold moderators offer a significant increase, up to a
factor of~10, in cold neutron brightness compared to a voluminous
moderator.  The obtained results are in excellent agreement with MCNP
calculations.

The chessboard or staircase assemblies of such moderators facilitate
the generation of wide neutron beams with simultaneously higher
brightness and intensity compared to a para-hydrogen-based cold
neutron source made of a single moderator~(either flat or voluminous) of the
same cross-section. Analytic model calculations indicate that gains
of up to approximately~2.5 in both brightness and intensity can be
achieved compared to a source made of a single moderator of
the same width.

The gain reduction in our study is mostly caused by these two factors: the limited
volume of the high-density thermal neutron region surrounding the
reactor core or spallation target, which restricts the total length of
the moderator assembly, and the finite width of moderator
walls. Additional factors affecting gain can be revealed through
dedicated Monte Carlo simulations of the moderator-reflector
assembly, which can only be conducted for a particular neutron source and
are beyond the scope of this general study. Moreover, the relatively
large length of moderator assemblies makes their application for short
pulse neutron sources very problematic.

The concept of ``low-dimensionality'' in moderators is explored,
demonstrating that achieving a substantial increase in brightness
necessitates moderators to be low-dimensional both geometrically,
implying a high aspect ratio, and physically, requiring the
moderator's smallest dimension to be smaller than the characteristic
scale of moderator medium (about the mean free path for thermal
neutrons). This explains why additional compression of the moderator
along the longest direction, effectively giving it a tube-like shape,
does not result in a significant brightness increase comparable to the
flattening of moderator.

\end{abstract}


\end{frontmatter}

\section{Introduction}
\label{sec:intro}

Cold neutron sources that provide a high cold neutron flux at neutron
scattering instruments are at the core of almost all neutron
scattering facilities. They exploit moderators made of a
hydrogen-containing material in liquid or solid state at cryogenic
temperatures~(e.g.\ hydrogen~\cite{kiyanagi2003},
deuterium~\cite{ageron1969}, methane~\cite{shin2010} and others), in
which the neutrons emitted from thermal moderators are thermalised
down to the temperature of a cold moderator. Recently, low-dimensional
para-hydrogen cold moderators were developed at the European Spallation
Source~(ESS)~\cite{Batkov2013a,Mezei2013a,Zanini2019a}. They employ a
remarkable feature of para-hydrogen: the drastic, more than an order of
magnitude difference in total neutron scattering cross-sections for
cold and thermal neutrons, that leads to the difference in their mean
free paths~(MFP), about \qty{10}{\cm} and \qty{1}{\cm},
respectively. This allows to squeeze usually large, about~\TenTenTen,
voluminous moderators to the flat shape~(so-called pancake
geometry~\cite{Batkov2013b}), with a height of about \qty{3}{\cm}. This
size is enough for thermal neutrons to be effectively slowed down in a
single collision to become cold neutrons, which can be extracted along
the whole large dimension of the moderator with a low parasitic
neutron absorption from para-hydrogen itself~\cite{Mezei2013a}. Indeed,
the brightness of the neutron beam emitted in this direction is
significantly, \numrange{2}{3} times enhanced in comparison to
the brightness of conventional voluminous moderators. Low-dimensional
para-hydrogen moderators are now getting more and more attention and are
considered for many newly designed high power and compact neutron
sources~\cite{SNSSTS, HBS, moskvin2023}. Being used with carefully
designed neutron transport systems~\cite{Konik2023a,Andersen2018},
such moderators can provide \numrange{2}{3} times gains in flux at
samples of neutron instruments requiring well-collimated incident
neutron beams.

In \secref{sec:2} we consider in details the mechanism leading to the
increase in brightness of flattened moderators. As thermal to cold
neutron conversion in single collision is a unique feature of
para-hydrogen~(discussed in details in \secref{sec:4}), we develop an
analytic model describing the moderators operating exclusively this
way. The model predicts significant brightness gains for a moderator
with high-aspect ratio viewed surface. The results are confirmed by
the corresponding MCNP~\cite{goorley2012} simulations.

In \secref{sec:3} we discuss the concept of ``low-dimensionality''
applied to neutron moderators.  We demonstrate that a neutron
moderator is low-dimensional and provides significant brightness gain,
only if it satisfies both criteria: geometrical
low-dimensionality~(high aspect ratio) of the viewed surface and
physical low-dimensionality, meaning at least one of moderator
dimensions is less than the characteristic scale of moderator
medium~(double MFP for thermal neutrons). Using this concept, we
explain why the squeezing of flat moderator to square shape does not
provide meaningful brightness gain, as found in~\cite{Mezei2013a}.

In \secref{sec:4} we use extensive MCNP calculations to investigate a
realistic para-hydrogen moderator and compare it to the previously developed
analytic model. We consequently complicate the model by adding
realistic illumination by Maxwellian thermal neutron spectrum and
actual para-hydrogen scattering cross-section, allowing for multiple
collisions process of cold neutron conversion.  The qualitative
behavior of brightness gains curves for real para-hydrogen is similar to one
predicted by the analytic model from \secref{sec:2} and one can still
observe high brightness gains.

In \secref{sec:5} we briefly overview some properties of the cold neutron spectrum emitted from low-dimensional para-hydrogen moderators.

In \secref{sec:6} we discuss trade-offs between brightness and
intensity of low-dimensional moderators. A~significant increase in
brightness of the emitted cold neutron beam is achieved at the cost of
a significant reduction in beam intensity. This fact seriously limits
the gain in the performance of the intensity-demanding instruments
with a low $Q$-resolution. Moreover, the use of moderators of small
size leads to the under-illumination of the entrance of a large
neutron guide that in turn results in a very irregular angular
structure of neutron beams at samples of high $Q$-resolution
instruments thus worsening their
performance~\cite{mattauch2017}. Therefore, it is an imperative to
find a possibility to increase the neutron beam intensity while
keeping a high brightness of the low dimensional moderators. We
propose employing chessboard and staircase  assemblies of narrow
moderators that provide wide, intense neutron beams while maintaining
the high brightness of the narrow moderator.

We also briefly consider the application of such moderators in the
case of their inhomogeneous thermal neutron illumination from a
nuclear reactor core or a spallation neutron source target. In
these cases, the stacked staircase~(chessboard) geometry allows for
simultaneous gains in brightness and intensity of up to
approximately~\num{2.5} for reactor sources and~\num{2} for spallation
or compact sources, relative to single flat moderators of the same
width. However, these gains will be affected by details of the
moderator-reflector assembly and should be estimated through dedicated
Monte Carlo simulations, which can only be conducted for a particular
neutron source and are beyond the scope of this general study.

\section{Analytic model: infinitely thin para-hydrogen moderator}
\label{sec:2}

\subsection{Brightness calculations for in-plane illumination}
\begin{figure}[b!]
\centering
\begin{subfigure}[t]{\whencolumns{0.7}{1}\linewidth}
\includegraphics[width=\linewidth]{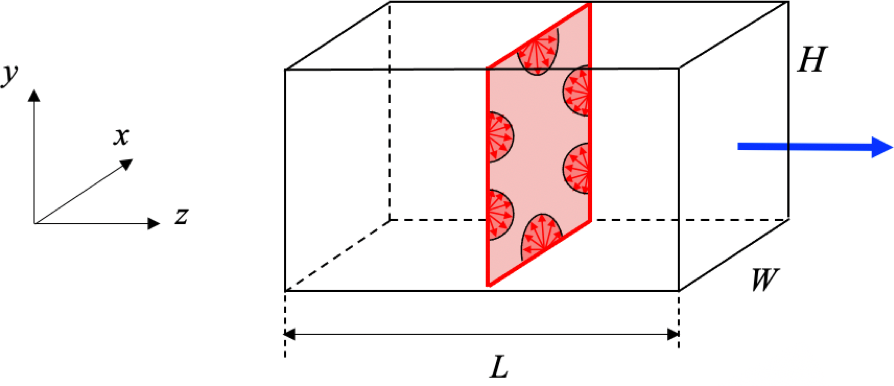}
\caption{}
\label{fig:slice:a}
\end{subfigure} \hfill
\begin{subfigure}[t]{\whencolumns{0.25}{1}\linewidth}
\includegraphics[width=\linewidth]{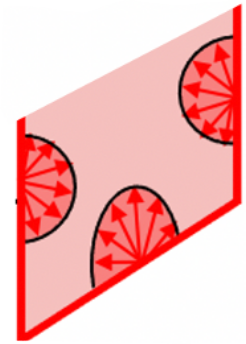}
\caption{}
\label{fig:slice:b}
\end{subfigure}\caption{(\subref{fig:slice:a})~Slice cut from the voluminous cold
neutron moderator of a rectangular parallelepiped \mbox{shape~($W
\times H \times L$)}. Red arrows indicate the directions of
thermal neutron illumination, also shown in detail
in~(\subref{fig:slice:b}). The blue arrow shows the direction of
the outgoing cold neutron beam towards a neutron optical
system~(e.g., neutron guide).}
\label{fig:slice}
\end{figure}

To calculate the brightness of a para-hydrogen moderator homogeneously
illuminated by thermal neutrons, we start with a simplified model of
an isolated, infinitely thin moderator slice cut from the voluminous
moderator. This slice is illuminated by thermal neutrons originating
from its perimeter~(as depicted in \figref{fig:slice}).

Initially, we focus on the case where each point of the linear thermal
neutron source~(represented by red arrows in \figref{fig:slice}) emits
neutrons isotropically within the $xy$-plane. We calculate the cold
neutron brightness in the $z$-direction under the assumption that
already a single collision with a hydrogen molecule is sufficient to
slow down thermal neutrons to cold neutron energies. The validity of
this assumption will be verified later in \secref{sec:4:1}, where we
analyse the energy losses of neutrons through multiple collisions
within the moderator.

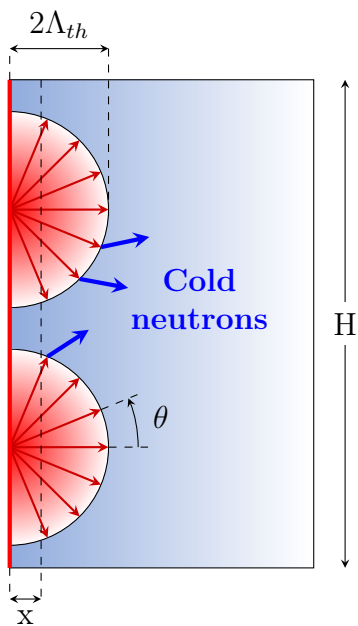
\begin{figure}
\centering
\tikzsetnextfilename{main-figure0}
\begin{tikzpicture}
\end{tikzpicture}

\caption{Para-hydrogen moderator illuminated by thermal neutrons
emitted from the side wall.}
\label{fig:2}
\end{figure}

First, we consider a para-hydrogen moderator illuminated by thermal neutrons
from one side, so that the side wall can be regarded as an isotropic
neutron source and investigate where and how cold neutrons are
produced. For this reason we examine the near-the-surface layer of
thickness $x$~(\figref{fig:2}). Thermal neutrons are propagating over
path $s$, that length depends on both the layer thickness $x$ and the
emission angle $\theta$ as $s = x/cos(\theta)$.

If a source element $dh$ emits neutrons within an angular range
$d\theta$, then the beam’s intensity can be expressed as follows:

\begin{equation}
\label{eq:1}
dI_{th0} = B_{th0} \, dh \, d\theta,
\end{equation}
where $B_{th0}$ is the brightness of a homogeneous isotropic linear
source.

The neutron
transmission after the propagation over the path $s$ is equal to
$\exp(-s/\Lambda_{th})$, where $\Lambda_{th}$ is the mean free path
of thermal neutrons in para-hydrogen, so that the intensity of the thermal
neutron beam is equal to

\begin{equation}
\label{eq:2}
dI_{th}(x,\theta) = dI_{th0}\exp\left(-\frac{x}{\Lambda_{th}\,\cos\theta}\right),
\end{equation}
where the initial intensity of the thermal beam is given in \eqref{eq:1}.

The intensity $dI_{cold}(x,\theta)$ of the beam comprised of neutrons
that have undergone collisions with the hydrogen molecules is defined
by reduction in thermal neutron beam intensity:

\begin{equation}
\begin{split}
\label{eq:3}
dI_{cold}(x,\theta) & = dI_{th0}-dI_{th}(x,\theta)= \\
& = dI_{th0}\left(1-\exp\Bigr(-\frac{x}{\Lambda_{th}\,\cos\theta}\Bigr)\right)= \\
& = B_{th0}\left(1-\exp\Bigr(-\frac{x}{\Lambda_{th}\,\cos\theta}\Bigr)\right) \, dh \, d\theta.
\end{split}
\end{equation}

Since we assumed that thermal neutrons are slowed down to cold
energies already as the result of a single collision~(see
\secref{sec:4:3} and \secref{sec:4:4} for more details),
$dI_{cold}(x,\theta)$ represents the intensity of the cold neutron
beam produced within a layer of thickness $x$ from the thermal
neutrons emitted with the initial direction $\theta$ by a perimeter
element $dh$. Total intensity $I_{cold}$ of the cold neutron beam
produced within such layer is obtained by integrating~\eqref{eq:3}
over $\theta$ and $h$:

\begin{equation}
\label{eq:4}
I_{cold}(x) = B_{th0}\, \int\limits_{0}^{H} \int\limits_{-\pi/2}^{\pi/2} \left(1-\exp\Bigl(-\frac{x}{\Lambda_{th}\,\cos\theta}\Bigr)\right) \, dh \, d\theta
\end{equation}

Here for simplicity we neglect the edge effects at the very top and
bottom of the moderator~(\figref{fig:2}), where thermal neutrons with
some directions $\theta$ can’t contribute to the production of cold
neutrons. However, these effects will be taken into account in further
calculations in \secref{sec:2:rec}.

The brightness of such cold neutron source is determined by dividing
intensity of the cold neutron beam by the size of the source:

\begin{equation}
\label{eq:Bcold}
B_{cold}(x) = \frac{I_{cold}(x)}{4 \pi H x}.
\end{equation}

\begin{figure}
\centering
\begin{subfigure}{\figwidth}
\tikzsetnextfilename{main-figure1}
\begin{tikzpicture}
\end{tikzpicture}
\caption{Cold neutron beam intensity}
\label{fig:3:a}
\end{subfigure}
\hspace{3em}
\begin{subfigure}{\figwidth}
\tikzsetnextfilename{main-figure2}
\begin{tikzpicture}
\end{tikzpicture}
\caption{Brightness gain}
\label{fig:3:b}
\end{subfigure}
\hspace{3em}
\caption{Intensity of the cold neutron beam~(a) and gain in
brightness~(b) as functions of the moderator layer thickness $x$
for thermal neutrons with mean free path $\Lambda_{th}
=\qty{1}{\cm}$~(blue curves) and $\Lambda_{th} = \qty{3}{\cm}$~(red
curves). The plots are normalised to the moderator of $x =
\qty{10}{\cm}$ thickness.}
\label{fig:3}
\end{figure}
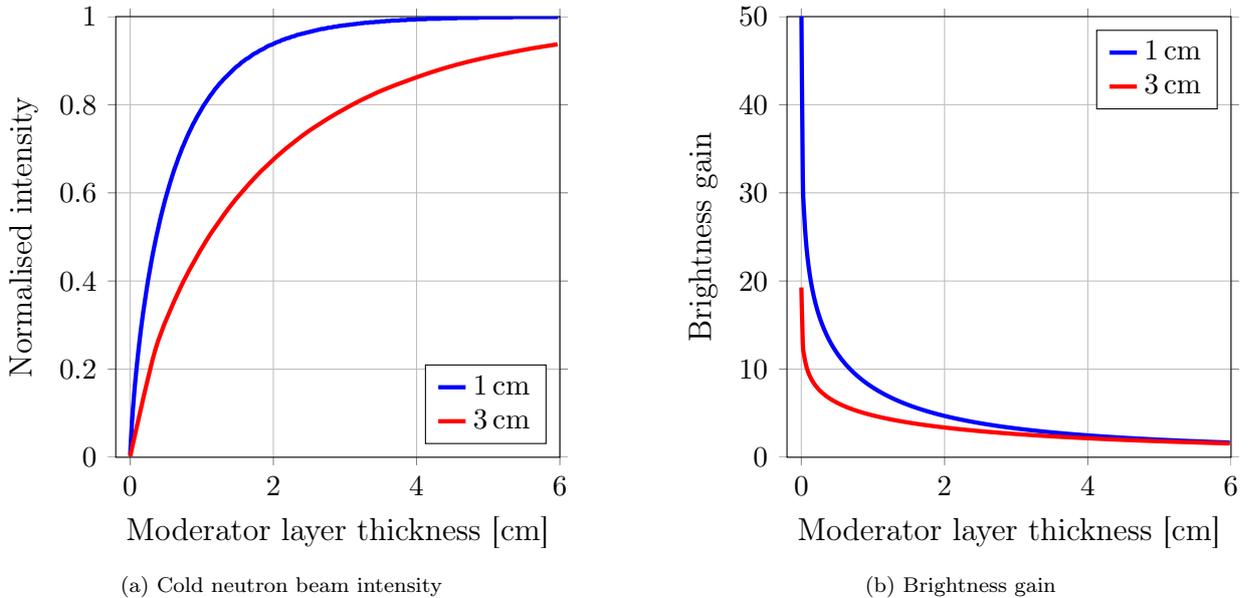

The dependence $I_{cold}(x)$ is shown in \figref{fig:3:a}. One can see
that about \qty{95}{\percent} of thermal neutrons are converted to cold
ones within the layer of $x=\TwoLambda$, so that there is no
significant increase in the cold neutron intensity for $x
>\TwoLambda$.  The red curve represents the behavior of neutrons with
a mean free path of $\Lambda_{th} = \qty{1}{\cm}$, while the blue curve
corresponds to neutrons with a mean free path of $\Lambda_{th} =
\qty{3}{\cm}$.  These mean free path values are associated with
a scattering cross-section $\sigma$ of \qty{23.9}{\barn} in para-hydrogen for
neutrons with an energy of \qty{60}{\milli\electronvolt}~(approximately
corresponding to the peak of the ESS water pre-moderator
spectrum~\cite[Figure 10]{Zanini2019a}) and
\qty{25}{\milli\electronvolt}, which is the most probable energy of
neutron spectra from H$_2$O or D$_2$O moderators in steady power
reactors at a temperature of \qty{300}{\kelvin}.

As the value of $x$ approaches zero, the numerator in \eqref{eq:Bcold} falls at a slower rate
than a linear function in denominator, so that the brightness
$B_{cold}(x)$ exhibits a significant increase~(\figref{fig:3:b}).

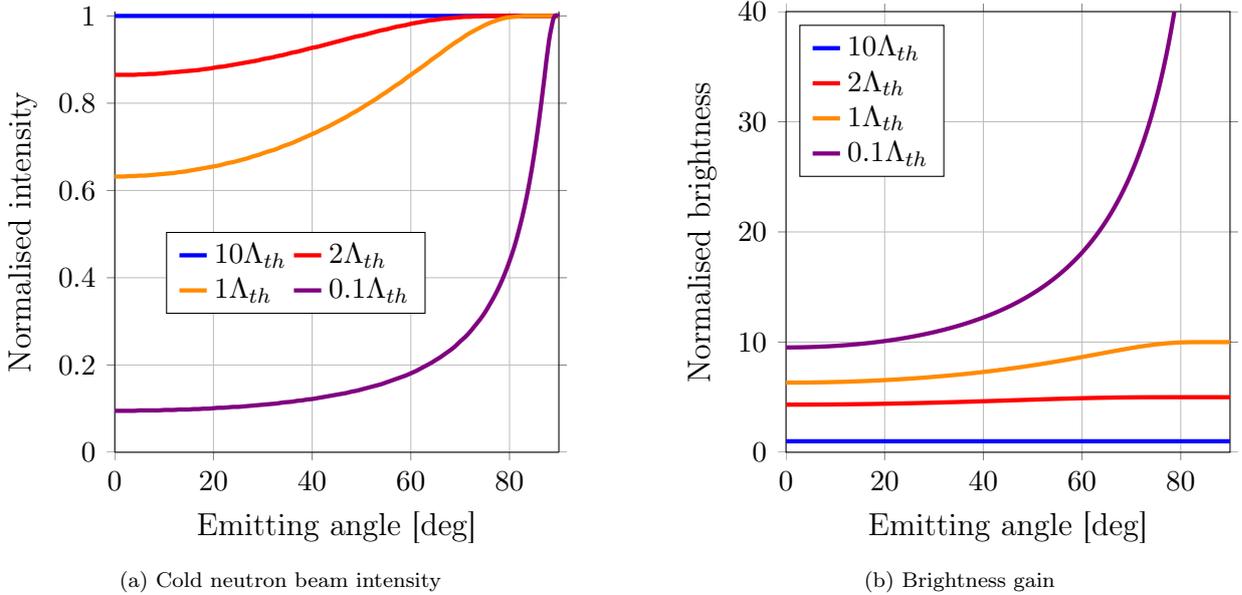
\begin{figure}
\centering
\begin{subfigure}{\figwidth}
\tikzsetnextfilename{main-figure3}
\begin{tikzpicture}
\end{tikzpicture}
\caption{Cold neutron beam intensity}
\label{fig:4:a}
\end{subfigure}
\hspace{3em}
\begin{subfigure}{\figwidth}
\tikzsetnextfilename{main-figure4}
\begin{tikzpicture}
\end{tikzpicture}
\caption{Brightness gain}
\label{fig:4:b}
\end{subfigure}
\caption{Intensity and brightness of the produced cold beam as
functions of the emitting angle~$\theta$ for different moderator
thicknesses.}
\label{fig:4}
\end{figure}

Let us consider cold moderator which width exactly matches the thickness $x$ of previously considered layer. Then we can conclude that
in the case of narrow moderators with a width $x \ll \Lambda_{th}$,
only ``grazing'' thermal neutrons, which propagate almost parallel to
the moderator wall with $\theta$ close to $\pm\pi/2$, have high
probability of colliding with hydrogen atoms and converting to cold
neutrons. This phenomenon is demonstrated by the angular dependence of
cold neutron intensity~(see \eqref{eq:3}) for different thicknesses of
the moderator layer~(\figref{fig:4:a}): the ``grazing'' neutrons
contribute significantly to the total cold neutron beam intensity,
leading to a substantial increase in brightness for narrow
moderators. This can be observed in \figref{fig:4:b} by comparing the
curves for $x=2\Lambda_{th}$, which roughly represents the thickness
of the flat cold moderator at ESS~\cite{Zanini2019a}, and
$x=0.1\,\Lambda_{th}$.

\subsection{Brightness gains calculations for rectangular moderator}
\label{sec:2:rec}

Using the approach described above, we can calculate the brightness of
the moderator when it is illuminated from all sides. We consider two
cases (\figref{fig:illumination}): illumination from the left and
right sides, and illumination from the top and bottom sides. In each
case, we can determine the partial cold neutron intensities by
calculating the corresponding neutron path lengths ($s_i$ and $b_i$,
respectively) for the four segments of the moderator area (shown
differently shaded in \figref{fig:illumination}) and for each neutron
emission point P defined by the positional offset $\Delta$ with
respect to the moderator center.  Total cold neutron intensities
$I_S(W,H)$ and $I_B(W,H)$ for the side and bottom illumination,
respectively, can be obtained by integrating over the moderator
height~$H$ or width~$W$ and the emission angle~$\theta$ measured from
the normal to the emitting surface. Detailed calculations of total
cold neutron intensities are provided in \ref{app:A}.

\begin{figure}
\centering
\begin{subfigure}[B]{\figwidth}
\hspace*{0.2\linewidth}
\tikzsetnextfilename{main-figure5}
\begin{tikzpicture}
\end{tikzpicture}
\caption{Side wall illumination}
\ifapp
\label{fig:app:illumination:a}
\else
\label{fig:illumination:a}
\fi
\end{subfigure}
\begin{subfigure}[B]{\figwidth}
\hspace*{0.2\linewidth}
\tikzsetnextfilename{main-figure6}
\begin{tikzpicture}
\end{tikzpicture}
\caption{Bottom wall illumination}
\ifapp
\label{fig:app:illumination:b}
\else
\label{fig:illumination:b}
\fi
\end{subfigure}
\ifapp
\caption{Notations used for calculating partial cold neutron
intensities. Note that these diagrams do not depict angles
$\theta_3$ and $\theta_4$.}
\label{fig:app:illumination}
\else
\caption{Illustration of infinitely thin moderator illumination}
\label{fig:illumination}
\fi
\end{figure}

Since the moderator is symmetric, the intensity $I_S(W,H)$~(refer to
equations \ref{eq:a:1}--\ref{eq:a:11}) is the same for both the
left and right walls. Similarly, the intensity $I_B(W,H)$~(refer to
equations \ref{eq:a:12}--\ref{eq:a:22}) is the same for both the
bottom and top walls.

From this point forward, we will calculate the brightness gain $G$
relative to the brightness $B(10,10) = I(10,10)/(4\pi\cdot\TenByTenCm)$ of the voluminous moderator with a
cross-section of \TenByTenCm~(that is the standard size of voluminous
para-hydrogen moderators~\cite{Ikeda2009}) under the same illumination
conditions:

\begin{eqnarray}
G_S(W,H) & = & \frac{B_S(W,H)}{B(10,10)} = \frac{I_S(W,H)}{4\pi \, (W \times H)} \cdot \left(\frac{I(10,10)}{4\pi \cdot \TenByTenCm}\right)^{-1}
\label{eq:gain:a} \\
G_B(W,H) & = & \frac{B_B(W,H)}{B(10,10)} = \frac{I_B(W,H)}{4\pi \, (W \times H)} \cdot \left(\frac{I(10,10)}{4 \pi \cdot \TenByTenCm}\right)^{-1}
\label{eq:gain:b}
\end{eqnarray}

As the cold neutron intensity delivered from the moderator is
proportional to the brightness of a homogeneous isotropic linear
source $B_{th0}$ that illuminates the moderator, brightness
gains~(\ref{eq:gain:a}) and~(\ref{eq:gain:b}) do not depend on
$B_{th0}$. This implies that, at a first-order approximation, any
inhomogeneity in the incident thermal neutron flux, such as that
arising from an unilluminated wall, does not affect the calculation of
brightness gain.

\begin{figure}
\centering
\begin{subfigure}[T]{0.1\textwidth}
\tikzsetnextfilename{main-figure7}
\begin{tikzpicture}
\end{tikzpicture}
\end{subfigure}
\hspace{1cm}
\begin{subfigure}[T]{0.4\textwidth}
\tikzsetnextfilename{main-figure8}
\begin{tikzpicture}
\end{tikzpicture}
\caption{Side wall illumination, various heights}
\label{fig:6:side:width}
\end{subfigure}
\begin{subfigure}[T]{0.4\textwidth}
\tikzsetnextfilename{main-figure9}
\begin{tikzpicture}
\end{tikzpicture}
\caption{Side wall illumination, various widths}
\label{fig:6:side:height}
\end{subfigure}

\begin{subfigure}[T]{0.1\textwidth}
\tikzsetnextfilename{main-figure10}
\begin{tikzpicture}
\end{tikzpicture}
\end{subfigure}
\hspace{1cm}
\begin{subfigure}[T]{0.4\textwidth}
\tikzsetnextfilename{main-figure11}
\begin{tikzpicture}
\end{tikzpicture}
\caption{Bottom wall illumination, various heights \newline (the curves are the same as in subfigure~\subref{fig:6:side:height})}
\label{fig:6:bottom:width}
\end{subfigure}
\begin{subfigure}[T]{0.4\textwidth}
\tikzsetnextfilename{main-figure12}
\begin{tikzpicture}
\end{tikzpicture}
\caption{Bottom wall illumination, various widths~(the curves are the same as in subfigure~\subref{fig:6:side:width})}
\label{fig:6:bottom:height}
\end{subfigure}
\caption{Moderator brightness gain as a function of its dimensions
for $\Lambda_{th} = \qty{1}{\cm}$ with various illumination
conditions.}
\label{fig:6}
\end{figure}

The results of the calculations for the partial brightness gains are
presented in \figref{fig:6}. It is evident from the graphs that these
brightness values exhibit a significant increase as the moderator
becomes narrower or as its height increases. This behavior can be
explained as follows:

\begin{itemize}
\item Side illumination~--- dependence on the moderator width $W$~(\figref{fig:6:side:width}).

As mentioned above, approximately \qty{95}{\percent} of thermal
neutrons interact with the moderator material within the \TwoLambda
thick layer~(see \figref{fig:3:a}), resulting in effective emission of
cold neutrons only from this region. Therefore, as the bottom width
$W$ of the moderator exceeds \TwoLambda, the intensity of cold
neutrons does not show a significant increase. In other words, for $W
> \TwoLambda$, the brightness of the cold source, which is inversely
proportional to the moderator area $W \times H$~(see
\eqref{eq:gain:a}), decreases as $1/W$ towards zero.

On the other hand, as the width $W$ decreases from \TwoLambda to
zero, the hyperbolic rise of $1/W$ for the cold neutron brightness is
dampened by the exponential decrease of the cold neutron intensity
within the \TwoLambda thick layer~(\eqref{eq:2}). As a result, the
brightness of cold neutrons increases rather sharply for narrower
moderators, but it does not diverge for $W = 0$; instead, it
approaches a limiting value.

\item Bottom illumination~--- dependence on the moderator width $W$~(\figref{fig:6:bottom:width}).

Unlike to the previous case, where the change in moderator width $W$
did not affect the size of the thermal neutron source~(vertical wall in \figref{fig:6}), in this case, it does have an
impact. For large values of $W$, more
than \TwoLambda, the cold neutron intensity is proportional to $W$,
as the vast majority of thermal neutrons emitted from each point on
the bottom wall are converted into cold neutrons in a single
collision. Therefore, the cold neutron brightness, as
defined by \eqref{eq:gain:b}, is independent of $W$ and inversely
proportional to the moderator height $H$.  However, as the width $W$
gradually decreases from \TwoLambda to zero, a portion of thermal
neutrons emitted at angles $\theta \neq 0$~(i.e.\ not vertically) will
not undergo collisions with the moderator material. This leads to a
decrease in the cold neutron intensity~(see \eqref{eq:4} and \figref{fig:3:a})

\item Side illumination~--- dependence on the moderator height $H$~(\figref{fig:6:side:height})

The dependence of cold neutron brightness on the moderator height for
the side illumination~(\figref{fig:6:side:height}) exhibits exactly
the same behavior as dependence on the moderator width for the
bottom illumination (\figref{fig:6:bottom:width}). Such kind of
symmetry is not surprising: in the first case, the illuminated
vertical side wall plays a similar role to the illuminated horizontal
wall in the case of bottom illumination.

\item Bottom illumination~--- dependence on the moderator height $H$~(\figref{fig:6:bottom:height}).

This dependence is but exactly the same as dependence on the
moderator width for the side
illumination~(\figref{fig:6:side:width}). Now the illuminated
horizontal bottom wall plays the same role as the illuminated vertical
wall for the side illumination.
\end{itemize}

The total moderator brightness is obtained as the sum of
left-right, $B_S(W,H)$~, and top-bottom, $B_B(W,H)$ partial brightness,

\begin{equation}
\label{eq:Bfull}
B (W,H) = B_S (W,H) + B_B (W,H),
\end{equation}

so that the brightness gain for fully
illuminated moderator is given by

\begin{equation}
\label{eq:Gfull}
G (W,H) = \dfrac{B(W,H)}{B(10,10)} = \dfrac{G_S (W,H) + G_B (W,H)}{2}.
\end{equation}

The factor of 2 in \eqref{eq:Gfull}here arises from the fact
that both brightness gains $G_S$ and $G_B$ are already normalized to
the brightness $B(10,10)$ of the voluminous moderator.

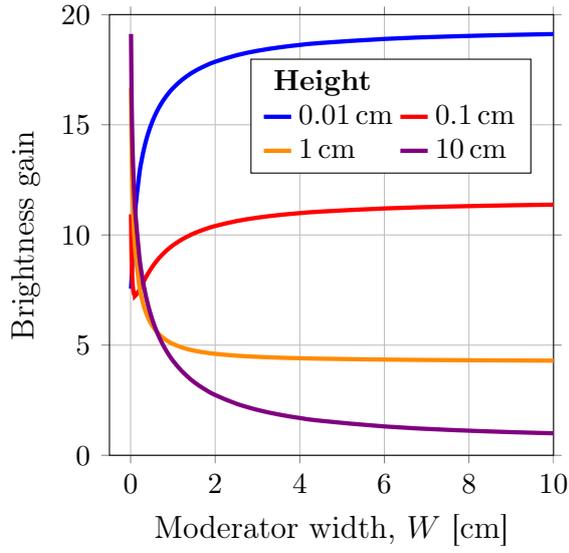
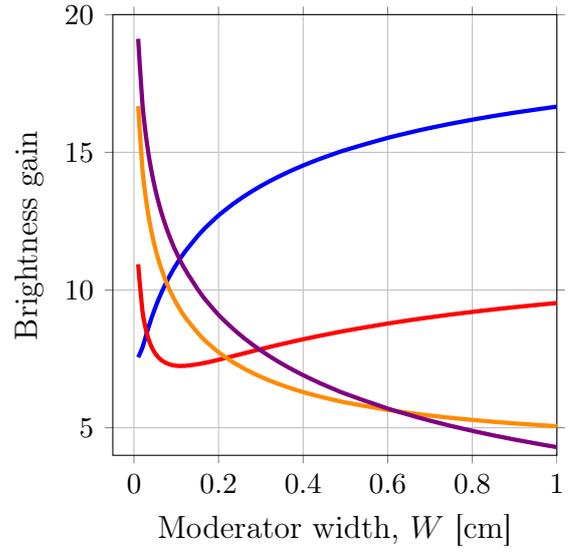
\begin{figure}
\begin{subfigure}{\figwidth}
\tikzsetnextfilename{main-figure13}
\begin{tikzpicture}
\end{tikzpicture}
\caption{Coarse scale}
\label{fig:7:a}
\end{subfigure}
\hspace{3em}
\begin{subfigure}{\figwidth}
\tikzsetnextfilename{main-figure14}
\begin{tikzpicture}
\end{tikzpicture}
\caption{Fine scale}
\label{fig:7:b}
\end{subfigure}
\caption{Brightness gain of fully illuminated moderator for different moderator heights}
\label{fig:7}
\end{figure}

The brightness gain of fully illuminated moderator is presented in
\figref{fig:7}~(for coarse~(\figref{fig:7:a}) and fine~(\figref{fig:7:b})
scales). One can see that rather high gains, about 10, can be
achieved for moderators with sizes of about \ZeroOneByTenCm.  Here the
minimal moderator width considered is $W =
\qty[round-precision=2]{0.01}{\cm}$, while for smaller values of $W$
even higher brightness gain can be observed. Note the non-monotonic
behavior of brightness for moderator of \qty{0.1}{\cm} height; the
physical explanation of this fact will be given in \secref{sec:3}.

As one can see from \figref{fig:8} below, there is a strong dependence of
the neutron cross-section~(or MFP) on the neutron energy in the region
of the maximum of the thermal Maxwellian spectrum. However, as it is
seen from \figref{fig:3}, the brightness gain is larger for shorter
MFPs, while characteristic features of the behavior of the brightness
gain are conserved.

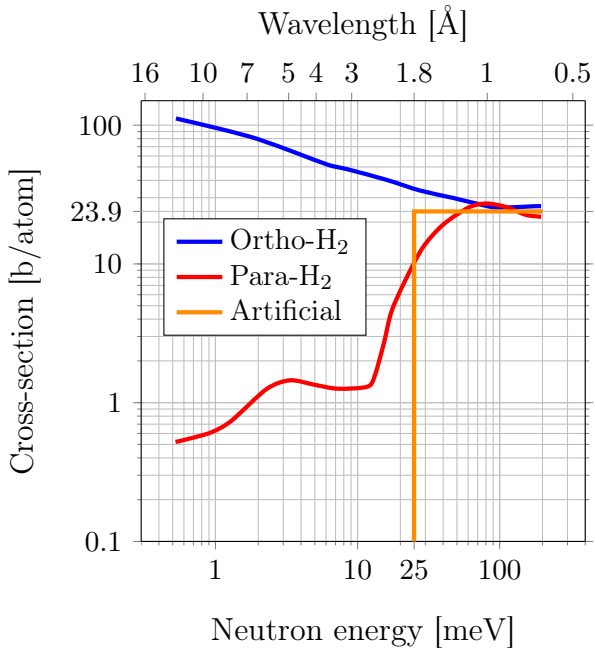
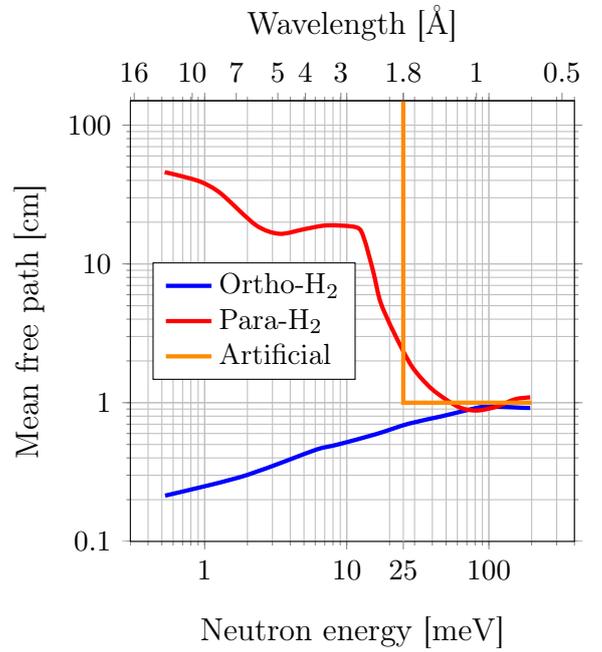
\begin{figure}
\centering
\begin{subfigure}[T]{\figwidth}
\tikzsetnextfilename{main-figure15}
\begin{tikzpicture}
\end{tikzpicture}
\caption{Cross-section.}
\label{fig:8:a}
\end{subfigure} \hspace{3em}
\begin{subfigure}[T]{\figwidth}
\tikzsetnextfilename{main-figure16}
\begin{tikzpicture}
\end{tikzpicture}
\caption{Mean free path.}
\label{fig:8:b}
\end{subfigure}
\caption{Energy dependence of cross-section and mean free path of
neutrons in ortho- and para-hydrogen at~\qty{20}{\kelvin}. Orange lines
correspond to the artificial step-like cross-section. Recalculated
from~\cite[figure 30 on page 367]{Watanabe2003}.}
\label{fig:8}
\end{figure}

It is important to note that the simple summing up of the two partial
brightness in~\eqref{eq:Bfull} is valid only if the cold moderator is
homogeneously illuminated from all sides. This assumption holds true
in the case of heavy water thermal moderators commonly used in
research reactors. However, if the cold moderator is placed in a light
water moderator or in close proximity to the target of an
accelerator-driven neutron source surrounded by reflectors, the
illumination becomes non-homogeneous. In such cases, the summing up of
the partial brightness needs to be performed with corresponding
weights to account for the non-uniform illumination.

It is worth mentioning that when only one side wall of the cold
moderator is illuminated the brightness gain is twice as much compared
to full illumination~(compare maximal gains in figures~\ref{fig:6} and
\ref{fig:7}).  However, these larger relative gains do not mean larger
absolute values of cold neutron brightness.

\subsection{MCNP verification}

The scattering of cold neutrons in para-hydrogen is very weak: the
microscopic scattering cross-section $\sigma_s$ of para-hydrogen for neutrons
with energies below \qty{13}{\meV} is only about \qty{1}{\barn}~(see
\figref{fig:8:a}) that corresponds to the mean free path
$\Lambda_{th}$ of about \qty{20}{\cm}. Therefore, cold neutrons
produced in the depth of the moderator volume can propagate over
distance $l = \qtyrange[range-phrase={-}]{10}{15}{\cm}$ to the moderator face without significant
losses.

Analytic calculations shown above predict significant brightness gains
for narrow cold moderators. To validate and further explore these
findings, it is crucial to compare them with results obtained from
MCNP simulations. However, there is a difference in these two
approaches in calculating the intensity of the cold neutron beam. In
the analytical calculations, the intensity is determined under the
assumption that a single collision is sufficient to slow down thermal
neutrons to cold energies. On the other hand, in MCNP simulations, the
number of cold neutrons produced in the moderator from thermal
neutrons is directly calculated using the scattering cross-section
$\sigma(E_n)$ for para-hydrogen. This means that multiple collisions are not
excluded in the MCNP simulations.

To ensure a fair comparison between our analytic calculations and MCNP
simulations, we need to limit the simulations to single
collisions. For this purpose, we replace the real cross-section
$\sigma(E_n)$~(shown in \figref{fig:8:a}) with an artificial step-like
cross-section $\sigma_{art}(E_n)$ that has a value of $\sigma_{art}=
\qty{23.9}{\barn}$~($\Lambda_{th} \approx \qty{1}{\cm}$) for thermal
neutrons with energies $E_n \ge \qty{25}{\meV}$, and a value of $\sigma_{art}
= 0$ for neutrons with $E_n < \qty{25}{\meV}$~(represented by a solid orange
line in \figref{fig:8:a}). Additionally, we set the energy of incident
thermal neutrons to \qty{25}{\milli\electronvolt}, so that even a small
energy loss due to a collision will result in the conversion of
thermal neutrons to cold neutrons. The neutrons are detected by the next event estimator~(f5),
which is a point detector placed at a distance of $r=\qty{100}{\meter}$
from the moderator. The point detector records cold neutrons with maximal
energy set to \qty[round-precision=2]{24.99}{\meV} emitted within the solid angle
$\Delta\Omega=S/r^2$, where $S$ is the emission surface of the
moderator, given by $S=W\times H$. Since the step-like cross-section
$\sigma_{art}(E_n)$ equals to zero for neutrons with energies below
\qty{25}{\meV}, there is no absorption of cold neutrons in para-hydrogen after
their conversion from thermal neutrons. This approach allows us to
focus solely on the effect of single collisions on the conversion
process in the MCNP simulations, enabling a direct comparison with the
results obtained from the analytic calculations.

\begin{figure}
\begin{subfigure}{\figwidth}
\tikzsetnextfilename{main-figure17}
\begin{tikzpicture}
\end{tikzpicture}
\caption{}
\label{fig:9:a}
\end{subfigure}
\hspace{3em}
\begin{subfigure}{\figwidth}
\tikzsetnextfilename{main-figure18}
\begin{tikzpicture}
\end{tikzpicture}
\caption{}
\label{fig:9:b}
\end{subfigure}
\caption{(a)~Comparison of MCNP simulations~(circles) with analytic
calculations~(curves) in \figref{fig:7}; (b)~Zoomed-in view of
\Figref{fig:9:a} for moderator widths below \qty{1}{\cm}.}
\label{fig:9}
\end{figure}
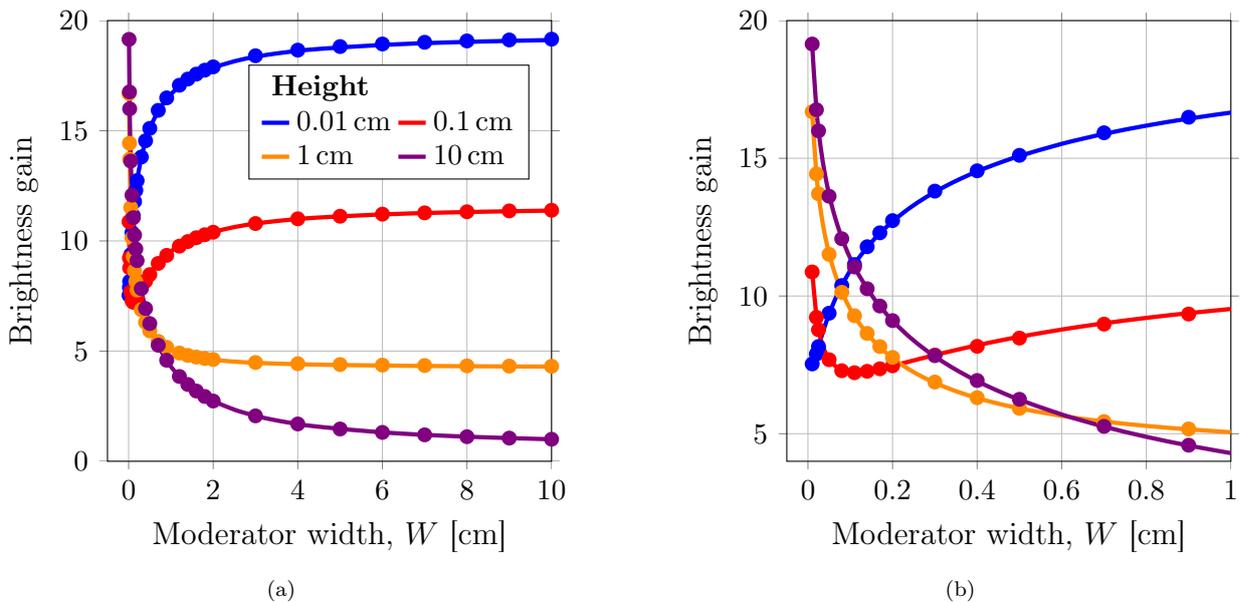

In \figref{fig:9} the results of the MCNP simulations are represented
by black symbols\footnote{The results of MCNP calculations are
consistently reported with statistical uncertainties. In cases where
the error bars are not discernible, it indicates that they fall within
the boundaries of the marker footprint.} and are superimposed on the
results obtained from the analytic calculations shown in
\figref{fig:7:a}. The excellent agreement between the two sets of
results demonstrates the accuracy and reliability of both the analytic
calculations and the MCNP simulations. This strong validation allows
us to confidently rely on the MCNP simulations for further analysis
and investigation of neutron moderation phenomena.

\section{What is the low-dimensionality of moderators?}
\label{sec:3}

Now we would like to discuss the physical question: what is the actual
meaning of ``low-dimensionality'' in moderators that leads to a
significant brightness gain?  In literature, moderators with
shapes such as tubes or disks that have face dimensions smaller than
the voluminous moderators' faces (typically about \TenByTenCm), are
commonly referred to as low-dimensional moderators
~\cite{Batkov2013a,Mezei2013a,Zanini2019a}. In the following
discussion, we will attempt to provide a physical interpretation of
the term ``low dimensionality''.

Low-dimensional moderators are created by compressing the voluminous
moderators in one or two directions. When considering only the
dimensions of the moderator faces~(height $H$ and width $W$),
classical voluminous moderators are geometrically 2-dimensional with
the aspect ratio $H/W$ of about~1. On other hand, the ESS-type flat
moderator~\cite{Zanini2019a} with an elongated face is geometrically
1-dimensional with an aspect ratio considerably larger than~1, and
provides significant brightness gain of about \numrange{2}{3} times
compared to voluminous moderator~(see the point at $W=\qty{3}{\cm}$ on
the violet curve in \figref{fig:9:a} for illustration).

Intuitively, one could assume further similar gains
of about \numrange{2}{3} times, which could be obtained by the
compression of moderator along its second dimension. However, MCNP
simulations have shown only modest increase of
\qtyrange{20}{25}{\percent}~(for neutron wavelength about
\qty{3}{\angstrom}) as shown in \cite[Figure 3]{Mezei2013a}.

We suggest the following explanation for this effect. Note, the
obtained tube-like moderator is geometrically 2-dimensional, since the
ratio $H/W$ is close to~1. Indeed, while the brightness gains are
rising monotonically in the sequence ``voluminous~$\to$ flat~$\to$
tube-like'' moderators, their dimensionality changes
non-monotonically: 2D~$\to$ 1D~$\to$ 2D. Thus, brightness gains can't
be explained solely from geometrical properties, i.e., the face
dimensionality. To be considered low-dimensional in terms of moderator
physics, in addition to having a high aspect ratio~(being
geometrically 1-dimensional), the moderator should also be
low-dimensional ``physically''. This physical dimension is defined by
the mean free path~$\Lambda_{th}$. More specifically, the physical
scale is~$2\Lambda_{th}$, which represents the distance beyond which
there is essentially no further production of cold neutrons, because
only a small number of thermal neutrons~(about \qty{5}{\percent}) can penetrate to
such depths in the moderator material.

\begin{figure}
\begin{subfigure}[t]{0.195\linewidth}
\centering
\tikzsetnextfilename{main-figure19}
\begin{tikzpicture}
\end{tikzpicture}
\caption{Very narrow (1D)}
\label{fig:classification:verynarrow}
\end{subfigure}
\begin{subfigure}[t]{0.195\linewidth}
\centering
\tikzsetnextfilename{main-figure20}
\begin{tikzpicture}
\end{tikzpicture}
\caption{Narrow (1D)}
\label{fig:classification:narrow}
\end{subfigure}
\begin{subfigure}[t]{0.195\linewidth}
\centering
\tikzsetnextfilename{main-figure21}
\begin{tikzpicture}
\end{tikzpicture}
\caption{Wide (2D)}
\end{subfigure}
\begin{subfigure}[t]{0.195\linewidth}
\centering
\tikzsetnextfilename{main-figure22}
\begin{tikzpicture}
\end{tikzpicture}
{\captionsetup{format=hang}\caption{1D geometrically, \\ 2D physically}}
\end{subfigure}
\begin{subfigure}[t]{0.195\linewidth}
\centering
\tikzsetnextfilename{main-figure23}
\begin{tikzpicture}
\end{tikzpicture}
{\captionsetup{format=hang}\caption{2D geometrically, \\ 1D~physically}}
\end{subfigure}
\caption{Classification of moderators according their geometrical
and physical dimensionality. High brightness gains are achieved
only by moderators that possess both a high aspect ratio~(1D
geometrically) and physical narrowness.}
\label{fig:classification}
\end{figure}
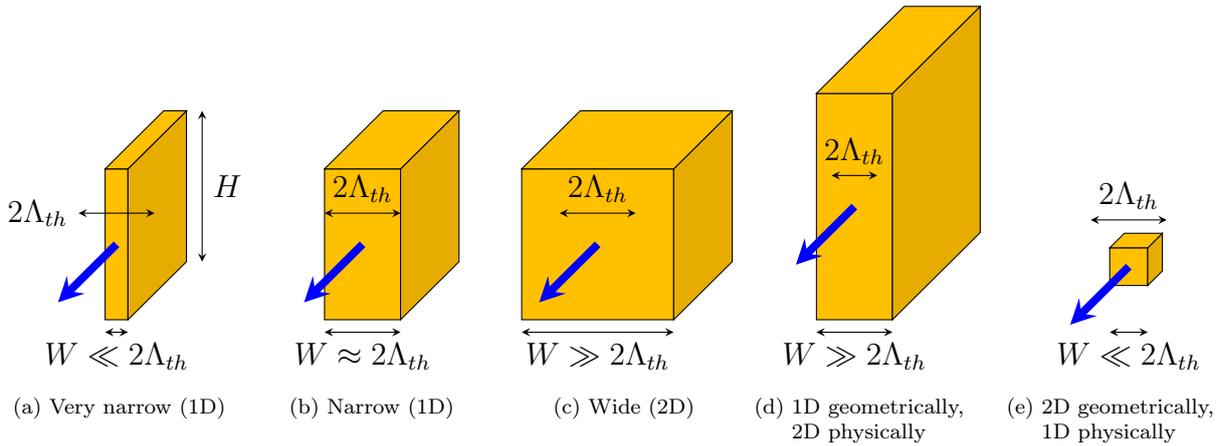

In our discussion, moderators with a width~$W \approx 2\Lambda_{th}$
are referred to as narrow moderators, while moderators with a width $W
\gg 2\Lambda_{th}$ are referred to as wide moderators.  Indeed, for a
moderator to be considered 1-dimensional, it needs to fulfill two
criteria: (a)~it should have a 1-dimensional shape, meaning it is
elongated in one direction and (b)~it should also be ``narrow'' with a
width $ W \approx 2\Lambda_{th} $.  Only moderators meeting both of
these criteria, as represented in
figures~\ref{fig:classification:verynarrow}
and~\subref{fig:classification:narrow}, are capable of delivering a
significant increase in brightness.

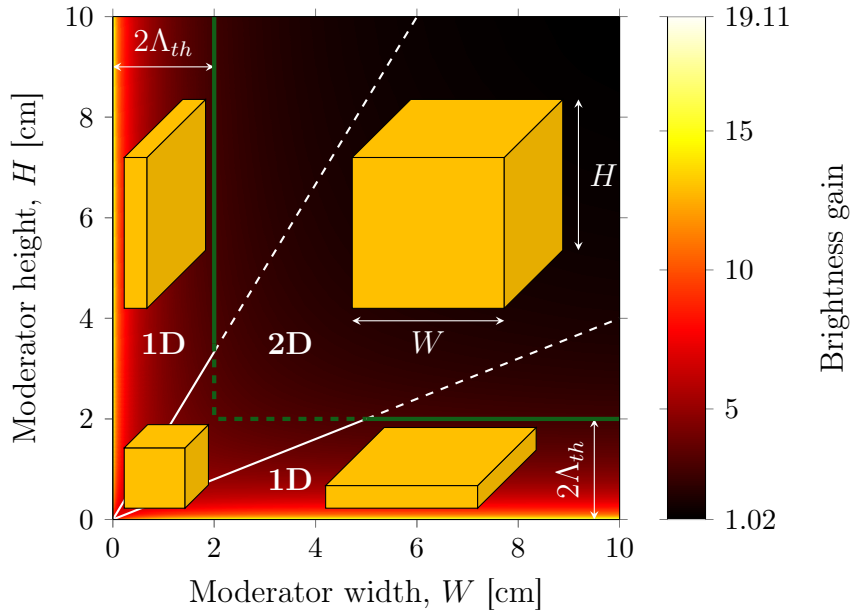
\begin{figure}
\centering
\tikzsetnextfilename{main-figure24}
\begin{tikzpicture}
\end{tikzpicture}
\caption{Brightness gain as a function of moderator width and
height. The figure also includes illustrations of different moderator
shapes associated with various regions of the map.}
\label{fig:map}
\end{figure}

This situation can be conveniently represented by a ``phase diagram''
shown in~\figref{fig:map}, which illustrates the relationship between
brightness gain and moderator width and height. Two inclined white
lines represent the boundaries where moderators have a 1-dimensional
geometric shape, characterised by a high aspect ratio. The green lines
represent the boundaries where the moderators are considered
1-dimensional physically, being narrow in width or height.

The areas enclosed by solid lines of both colours indicate regions
where high brightness gain is achieved, satisfying both geometric and
physical conditions. In contrast, the large central area represents
regions of low or no brightness gain.  The corresponding shapes of the
moderators for each area are also depicted in \figref{fig:map}.

The curves shown in \figref{fig:9} correspond to the horizontal
sections of \figref{fig:map}.  For example, the violet curve
corresponds to the cut at the very top of the diagram~($H =
\qty{10}{\cm}$): moderators with small $W$ provide significant
brightness gains, while moderator with $W = \qty{10}{\cm}$ provides
gain of exactly~1. Blue curve corresponds to the cut at the very
bottom of the diagram~($H = \qty[round-precision=2]{0.01}{\cm}$):
gains for moderators with small~$W$ are modest and become larger for
larger~$W$.

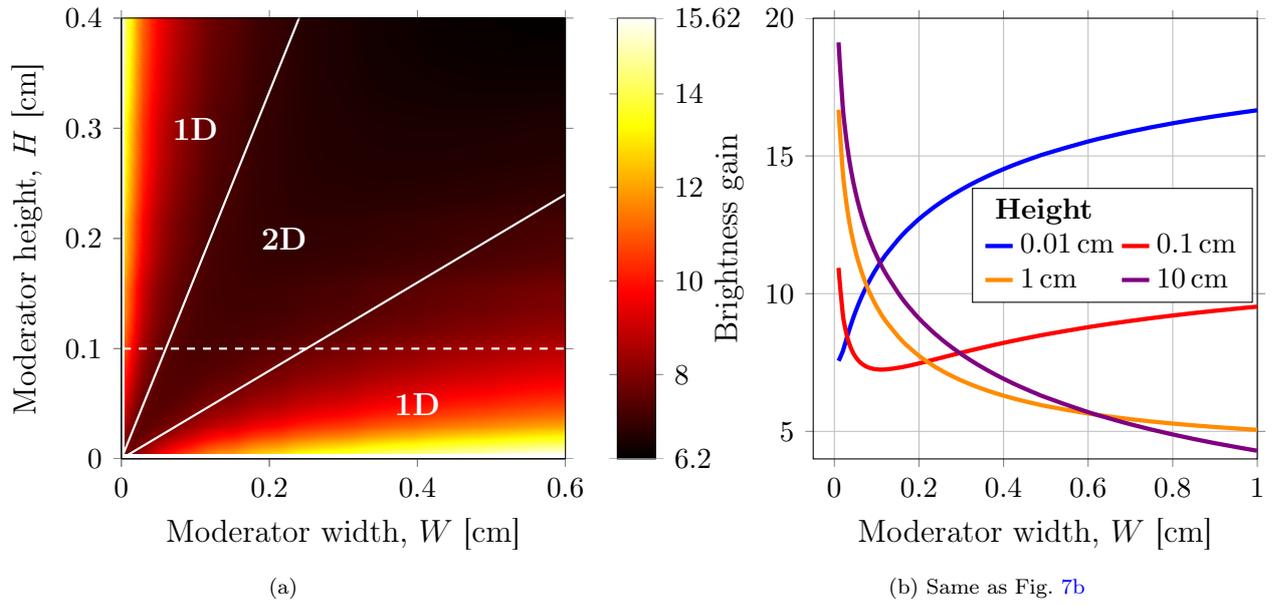
\begin{figure}
\begin{subfigure}[t]{\figwidth}
\tikzsetnextfilename{main-figure25}
\begin{tikzpicture}
\end{tikzpicture}
\caption{}
\label{fig:12:a}
\end{subfigure}
\hspace{4em}
\begin{subfigure}[t]{\figwidth}
\tikzsetnextfilename{main-figure26}
\begin{tikzpicture}
\end{tikzpicture}
\caption{Same as \figref{fig:7:b}}
\label{fig:12:b}
\end{subfigure}
\caption{(a)~Zoomed-in area of the region near the origin of
\figref{fig:map}. White dashed line corresponds to the
H=\qty{0.1}{\cm} curve in the \subref{fig:12:b}
subfigure. (b)~Brightness gain as a function of moderator width~(a
copy of \figref{fig:7:b}). Red curve corresponds to the cut
discussed in the text.}
\label{fig:12}
\end{figure}

This allows us to explain the non-monotonic dependence of brightness
gain for a moderator with height $H = \qty{0.1}{\cm}$ at small values
of width~$W$~(\figref{fig:9:b}). This effect is explored in
\figref{fig:12} where the zoomed-in area of the region near the origin
of \figref{fig:map} is shown. The change in the moderator width for a
fixed height value corresponds to the dashed line in
\figref{fig:12:a}, which represents a slice along $H = \qty{0.1}{\cm}$.
As the moderator width, $W$, increases, this line traverses through
1-dimensional, then 2-dimensional, and again 1-dimensional regions,
each associated with different brightness gains.

During this evolution, the shape
of the moderator changes from vertically elongated to horizontally
elongated~(see \figref{fig:map}); in both cases, the moderator exhibits a high brightness
gain. However, in between these two extremes, when the moderator has a
square shape, the brightness gain is lower as it is depicted in
\figref{fig:12:b}~(that is a copy of \figref{fig:7:b} and placed
here for the reader’s convenience).

\begin{figure}
\centering
\includegraphics[width=0.62\linewidth]{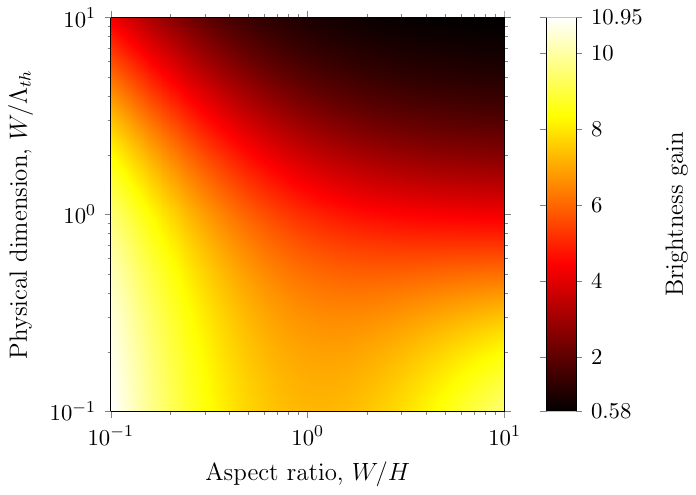}
\caption{Brightness gain as a function of physical dimension and aspect ratio.}
\label{fig:low_diagram}
\end{figure}

\Figref{fig:low_diagram} exhibits the same data as \figref{fig:map},
but plotted with respect to different coordinates. Here, instead of
moderator width and height we use ``physical dimension'', denoted
as~$W/\Lambda_{th}$, and ``geometrical dimension'', represented by the
aspect ratio~$W/H$. This representation visually supports our
conclusions about the physical meaning of low-dimensionality of
moderators. Generally, the brightness gains are higher for lower
physical dimension and high aspect ratio. If~$W$ is significantly
different from $H$, such that both $W/H \gg 1$ and $W/H \ll 1$ cases are
classified as ``high'' aspect ratios and correspond to horizontally or
vertically elongated moderators, respectively, this difference is
irrelevant for our considerations.

Note relatively low gains for aspect
ratio of unity, even for physically low-dimensional moderators.

The highest possible gains are achieved when both physical and geometrical
dimensions are low.

\section{Real para-hydrogen moderator}
\label{sec:4}

Up to now, we have considered a model of a para-hydrogen moderator with the
following assumptions: (a)~the moderator is infinitely thin, (b)~it is
illuminated by a linear thermal neutron source that emits neutrons
isotropically but only in the moderator plane, (c)~the thermal
neutrons are monochromatic with $E = \qty{25}{\meV}$, and (d)~the step-like
scattering cross-section, so that the conversion of thermal neutrons
to cold neutrons takes place in a single collision with a hydrogen
molecule.

In reality, moderators are of finite length and they are uniformly
illuminated by thermal neutrons with Maxwellian energy distribution
and the conversion to cold neutrons can involve multiple collisions~---
indeed, neither of the above-mentioned assumptions holds. Now, we will
proceed by consequently discarding these assumptions in order to
perform calculations involving a real moderator.

\subsection{Transition from infinitely thin to finite length moderator.}
\label{sec:4:1}

\begin{figure}
\begin{subfigure}[b]{\figwidth}
\includegraphics[width=0.9\linewidth]{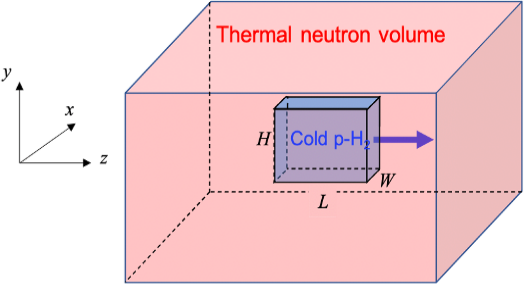}
\vspace{2cm}
\caption{}
\label{fig:13:a}
\end{subfigure}
\hspace{3em}
\begin{subfigure}[b]{\figwidth}
\tikzsetnextfilename{main-figure27}
\begin{tikzpicture}
\end{tikzpicture}
\caption{}
\label{fig:13:b}
\end{subfigure}
\caption{(a)~Para-hydrogen moderator in the thermal neutron volume created by
a large thermal moderator~(not shown). Blue arrow shows the direction of the emitted cold neutron beam. (b)~Brightness gain for the cold moderator illuminated from left, right, top and bottom surfaces as calculated with MCNP.}
\label{fig:13}
\end{figure}

Let's consider a para-hydrogen moderator that is immersed in a thermal
neutron volume created by thermal moderator, as shown in
\figref{fig:13:a}. In this configuration, all four surfaces of the
para-hydrogen moderator~--- left, right, top, and bottom (with the neglect of
illumination on the front and back faces)~--- are exposed to thermal
neutrons and indeed can be considered as emitters of thermal neutrons.

To analyze this
configuration, we can divide the para-hydrogen moderator into thin slices,
similar to what we discussed earlier. Each slice emits cold neutrons
along the length $L$ of the moderator towards the point detector. Since cold
neutrons are now generated in the volume $W\times H\times L$, the
brightness gain is calculated relative to the volume of the moderator,
which has dimensions of $(10 \times 10 \times L)$\,\unit{\cm\cubed},
with $L$ representing the length of the moderator.

\figref{fig:13:b} illustrates the results of MCNP
simulations where the thermal neutron
source is placed along the perimeter of an infinitely thin para-hydrogen
slices and neutrons are emitted isotropically in the~$(x,y)$ plane.
As expected, there is no difference between the brightness gain in
this configuration and that of a single slice, as shown in
\figref{fig:9}.

\subsection{Transition to out-of-plane illumination}
\label{sec:4:out}

Now we will extend our analysis to the case where each of infinitely
thin thermal neutron sources placed along the perimeter of each of the
para-hydrogen slices~(which altogether constitute the finite length
moderator) emits neutrons isotropically in space rather than in the
$(x,y)$ plane as it was considered above. Indeed, now the production
of cold neutrons from any particular thin thermal neutron source
occurs not solely within the corresponding thin slice but within a
broader layer with a thickness of about $4\Lambda_{th}$ in the
$z$-direction, where \qty{95}{\percent} of the cold neutrons are born.

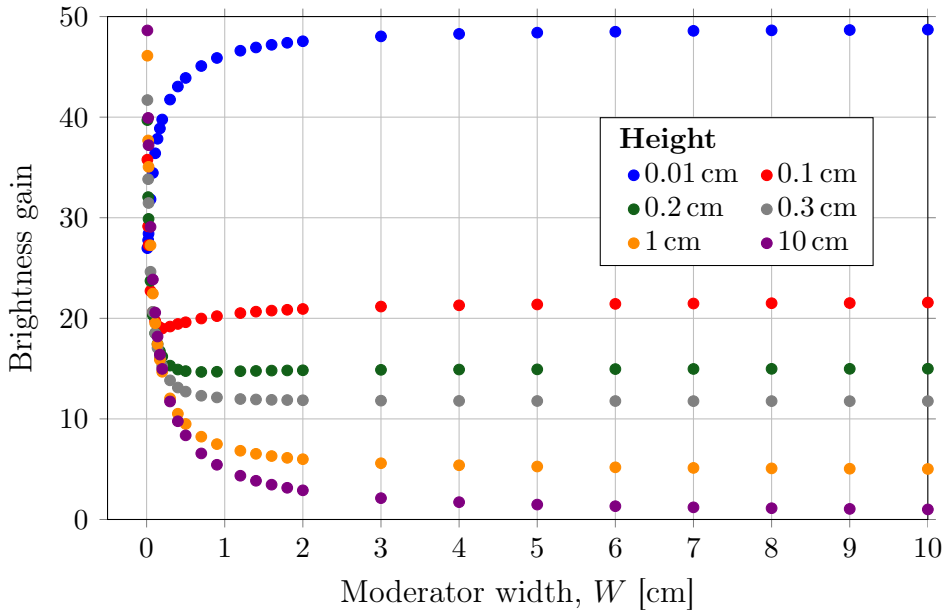
\begin{figure}
\centering
\tikzsetnextfilename{main-figure28}
\begin{tikzpicture}
\end{tikzpicture}
\caption{Brightness gains for the out-of-plane illumination.}
\label{fig:14}
\end{figure}

The dependence of brightness gains on the width and height of the
moderator for out-of-plane illumination is illustrated in
\figref{fig:14}. This figure demonstrates that the brightness gain is
approximately three times greater when compared to the results
obtained for in-plane illumination, as shown in \figref{fig:13:b}. To
understand the reasons for the greater brightness gain in the
out-of-plane case, it is important to note that the origin of thermal
neutrons~(the perimeter of the moderator slice in both cases) is not
relevant for calculating the brightness gain. Instead, what matters is
the location where these neutrons convert into cold neutrons.

\begin{figure}
\begin{subfigure}{\whencolumns{0.7}{1.0}\linewidth}
\includegraphics[width=\linewidth]{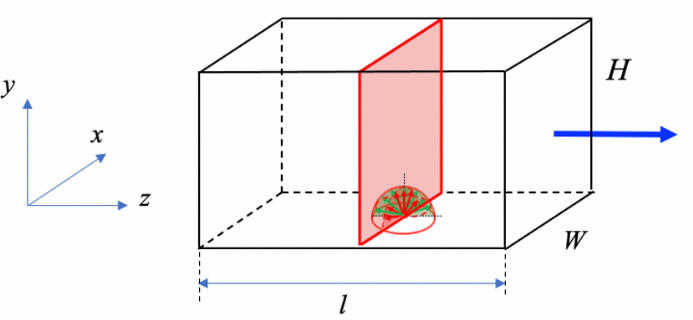}
\caption{}
\label{fig:15:a}
\end{subfigure}\hfill
\begin{subfigure}{\whencolumns{0.2}{1.0}\linewidth}
\centering
\includegraphics[width=\whencolumns{1.0}{0.5}\linewidth]{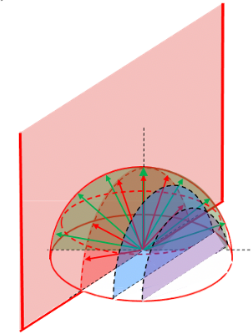}
\caption{}
\label{fig:15:b}
\end{subfigure}
\caption{Out-of-plane illumination of cold moderator by an
infinitely thin thermal neutron source placed along the
perimeter of the para-hydrogen slice. (a)~The red and green arrows indicate
the directions of thermal neutron illumination in the $(x, y)$
and $(y, z)$ planes, respectively; the blue arrow represents the
direction of the outgoing cold neutron beam towards the neutron
optical system, typically a neutron guide. (b)~The zoomed-in view: a hemisphere can be represented as a
combination of smaller semicircles normal to the $z$-axis and
offset from each along it}
\label{fig:15}
\end{figure}

Let's consider a simplified scenario where thermal neutrons are
emitted from a single point~(\figref{fig:15}). In the in-plane case,
the region where \qty{95}{\percent} of the cold neutrons are born can
be described as a semicircle with a radius of $2\Lambda_{th}$. In the
out-of-plane case, this region becomes a hemisphere with the same
radius and can be represented as a combination of smaller semicircles
normal to the $z$-axis and offset from each other along this axis~(as
shown in \figref{fig:15:b}). For these offset semicircles, the region
of cold neutron origination is compressed, effectively reducing the
mean free path of the thermal neutrons. However, as we already know,
the brightness gain increases with a decrease in the mean free
path~(as shown in \figref{fig:3:b}). Therefore, the gain will be
greater for all these offset semicircles compared to the in-plane
case. Since the overall brightness gain is the sum of the gains for
each of these semicircles, the total brightness gain in the
out-of-plane case is greater than in the in-plane case.

It is also possible to estimate the brightness gain analytically in
the out-of-plane case by following a similar approach as in the
two-dimensional case described in \secref{sec:2}. This involves applying
an additional integration over the out-of-plane angle.

\subsection{Transition from monochromatic to Maxwellian spectrum illumination }
\label{sec:4:3}

\begin{figure}
\centering
\tikzsetnextfilename{main-figure29}
\begin{tikzpicture}
\end{tikzpicture}
\caption{Brightness gains calculated for Maxwellian spectrum
illumination and artificial neutron scattering cross-section~(see
\figref{fig:8:a}).}
\label{fig:16}
\end{figure}
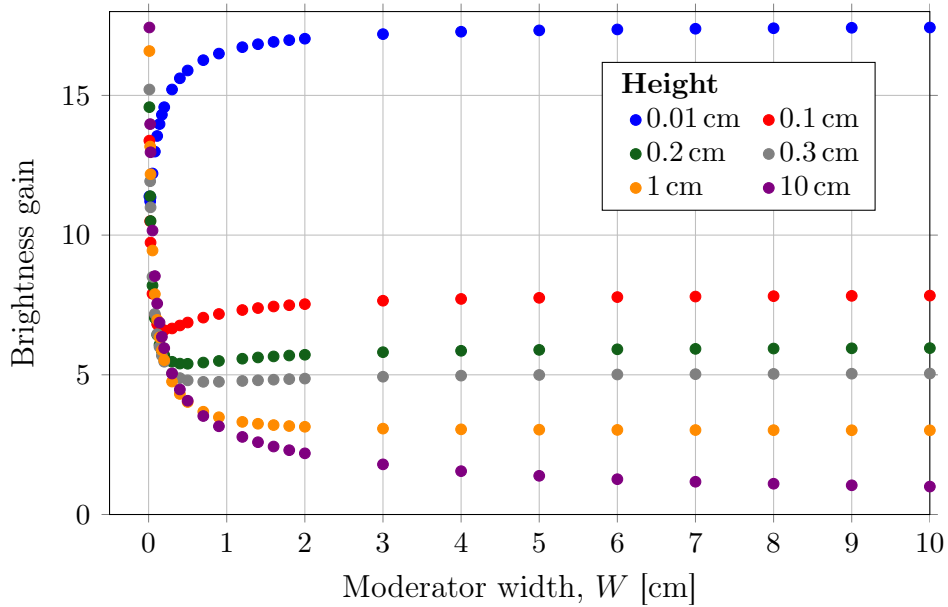

Now, we will shift from illuminating the moderator with a
monochromatic spectrum to using a Maxwellian spectrum with temperature
\qty{300}{\kelvin}. The results of the calculations are presented in \figref{fig:16}.  As
seen in this figure, when compared to \figref{fig:14}, we observe a
significant reduction in brightness gains, approximately by a factor
of~3, upon such replacement.

To explain this effect, we should note that up until now, our focus
has been on the process of slowing down thermal neutrons in the para-hydrogen
moderator, assuming a single collision as the sole mechanism. This was
achieved by utilizing an artificial step-like neutron scattering
cross-section with a threshold at \qty{25}{\milli\electronvolt}~(see \figref{fig:8}) and a
monochromatic neutron spectrum with the same energy. However, neutrons
with sufficiently high energies present in the incident Maxwellian
spectrum are, in principle, capable of undergoing two or more
collisions within the moderator before their energy decreases to less
than \qty{25}{\milli\electronvolt}, allowing them to exit.

\figref{fig:17} illustrates the breakdown of the data presented in
\figref{fig:16}, categorized by the number of collisions. It's
important to note that \figref{fig:17:a}, displaying the brightness
gains of cold neutrons generated through a one-collision process,
closely resembles \figref{fig:14}: this indicates minimal difference between
monochromatic and Maxwellian spectral illuminations in this particular
scenario.

\begin{figure}
\begin{subfigure}{\figwidth}
\tikzsetnextfilename{main-figure30}
\begin{tikzpicture}
\end{tikzpicture}
\caption{Single collision only}
\label{fig:17:a}
\end{subfigure}
\hspace{3em}
\begin{subfigure}{\figwidth}
\tikzsetnextfilename{main-figure31}
\begin{tikzpicture}
\end{tikzpicture}
\caption{Two and more collisions}
\label{fig:17:b}
\end{subfigure}
\caption{Decomposition of brightness gains shown in \figref{fig:16} on different number of collisions}
\label{fig:17}
\end{figure}

Now we explore \figref{fig:17:b}, where gains for multiple collision process
are shown in dependence of moderator width. Let's start by analyzing the curves that relate to
moderators with smaller heights. For the multiple collision process to
be effective, the para-hydrogen volume surrounding the initial collision site
must be significantly larger than a sphere with a radius of
$2\Lambda_{th}$, because after the first collision the slowed down
neutrons are scattered isotropically in all directions~(in $4\pi$). In
other words, the moderator dimensions in all directions should be much
greater than $2\Lambda_{th}$ to ensure sufficient volume for multiple
collision events.

Indeed, for narrow moderators with dimensions $W \ll H$ and $W \ll L$,
where $H$ and $L$ are much larger than $2\Lambda_{th}$, the
probability of a second collision event is relatively small. This is
because the second collision is mainly possible for neutrons that are
scattered along the large surfaces of the moderator, which corresponds
to a relatively small solid angle of about $4WL/H^2$. For this reason,
the brightness rapidly decreases as the width approaches zero.

On the other hand, in wide moderators where all~3 dimensions $W$, $H$, and
$L$ are much larger than $2\Lambda_{th}$, the probability of multiple
collisions~(two or more) is almost unity due to a large volume
available for neutron scattering and subsequent collisions. These can
be interpreted as a single quasi-collision with an increased mean free
path. The entire previously developed theory regarding brightness
enhancement with a reduction in moderator width remains applicable to
this quasi-collision scenario, predicting an increase in
brightness. This observation aligns with the violet curve shown in
\figref{fig:17:b}. However, at small widths, less than \qty{0.5}{\cm}, when two
collisions no longer remain assured, the scenario reverts to what was
described earlier, resulting in a decrease in brightness. Therefore,
the scattering of thermal neutrons with energies less than \qty{30}{\meV} on
para-hydrogen molecules is primarily a single collision process: these
neutrons experience a reduction of approximately \qty{15}{\meV} in their
kinetic energy during the initial collision. Subsequently, due to the
diminishing scattering cross-section for neutrons possessing energies
below \qty{15}{\meV}, the likelihood of undergoing a second collision becomes
negligible.

In this way, when illuminated by a Maxwellian spectrum, a large
\TenTenTen moderator is capable of producing cold neutrons through
both single-collision and multiple-collision processes. As the
moderator size decreases, the single-collision process leads to an
increase in brightness, while the multiple-collision process results
in a decrease. With an increase in the moderator's size, the
multiple-collision process begins to dominate, leading to an increase
in the absolute brightness of the large moderator. Consequently, this
leads to a reduction in the relative gain of the thin
moderator~(compare \figref{fig:14} and \figref{fig:16}).

The expected magnitude of this reduction can be estimated using the
results of MCNP calculations for the wavelength dependencies of the
brightnesses and the fractions of single and multiple
collisions. These calculations were performed for high-aspect ratio
\ZeroZeroOneByTenCm and voluminous \TenByTenCm moderators, and the outcomes
are presented in Figures~\ref{fig:18} and \ref{fig:19}, respectively.

\begin{figure}
\centering
\hspace{-1em}
\begin{subfigure}{\figwidth}
\tikzsetnextfilename{main-figure32}
\begin{tikzpicture}
\end{tikzpicture}
\caption{Brightness gains}
\label{fig:18:a}
\end{subfigure} \hspace{3em}
\begin{subfigure}{\figwidth}
\tikzsetnextfilename{main-figure33}
\begin{tikzpicture}
\end{tikzpicture}
\caption{Fraction of single and multiple collisions}
\label{fig:18:b}
\end{subfigure}
\caption{Wavelength dependence of brightness gains and fractions of
single and multiple collisions for voluminous \TenByTenCm moderator,
calculated for Maxwellian spectrum illumination and artificial
neutron scattering cross-section~(see \figref{fig:8:a}).}
\label{fig:18}
\end{figure}
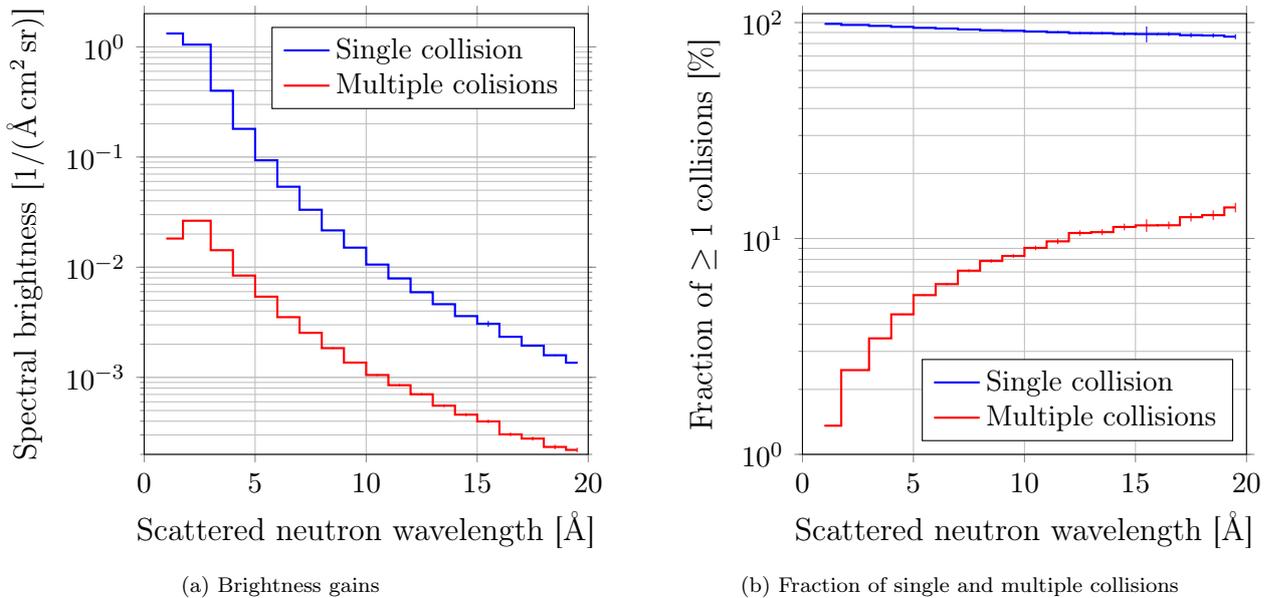

\begin{figure}
\centering
\begin{subfigure}{\figwidth}
\tikzsetnextfilename{main-figure34}
\begin{tikzpicture}
\end{tikzpicture}
\caption{Brightness gains}
\label{fig:19:a}
\end{subfigure} \hspace{3em}
\begin{subfigure}{\figwidth}
\tikzsetnextfilename{main-figure35}
\begin{tikzpicture}
\end{tikzpicture}
\caption{Fraction of single and multiple collisions}
\label{fig:19:b}
\end{subfigure}
\caption{Wavelength dependence of brightness gains and fractions of
single and multiple collisions for voluminous \ZeroZeroOneByTenCm
moderator, calculated for Maxwellian spectrum illumination and
artificial neutron scattering cross-section~(see
\figref{fig:8:a}).}
\label{fig:19}
\end{figure}

The reduction in gain factor is directly related to the share of
single-collision process in the total production of cold neutrons in
the voluminous moderator. This share can be deduced from
\figref{fig:19:b} by integrating over the whole spectrum with respect
to the spectral brightness~(\figref{fig:19:b}). Apparently this share
accounts for approximately \qty{25}{\percent} of all the cold
neutrons. From here, one can roughly estimate the brightness gain
reduction factor as~4, which would have led to a gain factor of
about~12.5 for the \ZeroZeroOneByTenCm moderator, instead of the
observed value of~17 (see \figref{fig:14} and
\figref{fig:16}). However, these considerations do not take into
account that some cold neutrons, produced by multiple collisions in
the voluminous moderator, are not completely lost with the decrease of
the moderator size. Instead, some of them undergo only single
collision in a smaller moderator and contribute to the final count of
single-collision produced neutron, thus increasing the gain factors
from~12.5 to~17.

\subsection{Transition from step-like cross-section to real cross-section of para-hydrogen }
\label{sec:4:4}

\begin{figure}
\centering
\tikzsetnextfilename{main-figure36}
\begin{tikzpicture}
\end{tikzpicture}
\caption{Brightness gains calculated for Maxwellian spectrum
illumination and real para-hydrogen cross-section.}
\label{fig:20}
\end{figure}
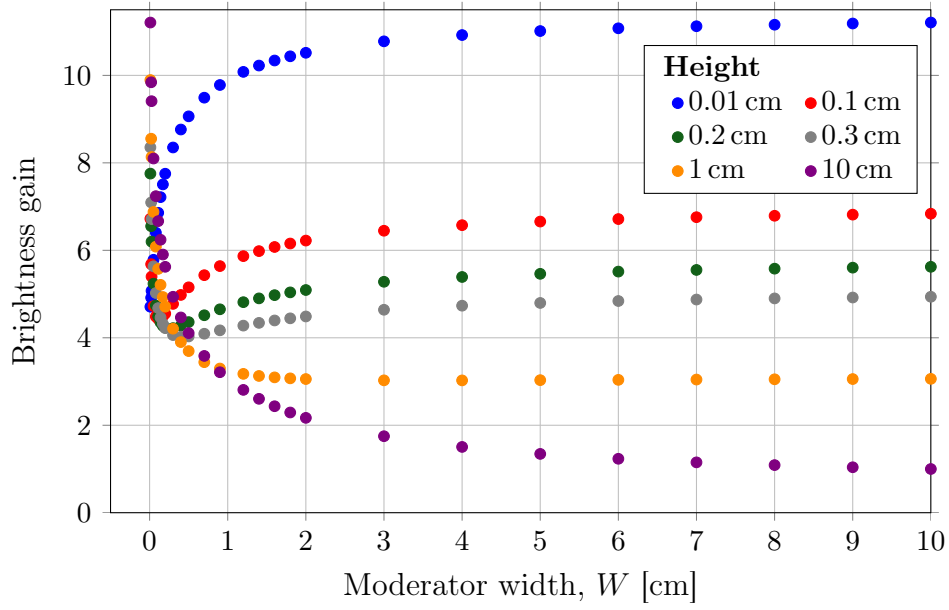

In the final phase, we replace the artificial neutron scattering
cross-section with a real para-hydrogen cross-section at
\qty{20}{\kelvin}~({\tt hpara.20t}~\cite{conlin2012,chadwick2006}). This substitution
results in an additional reduction in the brightness gain,
approximately by a factor of \num{1.5}, which varies depending on the
moderator width~(\figref{fig:20}).

To elucidate this effect, we will adopt a similar method as in the
preceding paragraph and break down the gains based on the number of
collisions, as shown in \figref{fig:21}. Comparing the gains obtained from
single collisions~(\figref{fig:21:a}) with those presented in \figref{fig:17:a}, we
can observe a reduction of approximately 2.5 times. This is mainly
attributed to the fact that in these processes neutrons with lower
energies are involved, for which the MFP at \qty{25}{\milli\electronvolt} in the previously
used step-like cross-section was \qty{1}{\cm}, while in the currently
used real cross-section the MFP is about \qty{2}{\cm}~(see
\figref{fig:8:b}). However, as demonstrated in analytical calculations~(see \figref{fig:3}),
for longer MFPs the relative brightness gains when
compressing the moderator are smaller, as observed here.

\begin{figure}
\begin{subfigure}{\figwidth}
\tikzsetnextfilename{main-figure37}
\begin{tikzpicture}
\end{tikzpicture}
\caption{Single collision only}
\label{fig:21:a}
\end{subfigure}
\hspace{3em}
\begin{subfigure}{\figwidth}
\tikzsetnextfilename{main-figure38}
\begin{tikzpicture}
\end{tikzpicture}
\caption{Two and more collisions}
\label{fig:21:b}
\end{subfigure}
\caption{Decomposition of brightness gains shown in \figref{fig:20} on different number of collisions}
\label{fig:21}
\end{figure}

Regarding the multiple-collision processes~(see \figref{fig:17:b} and
\ref{fig:21:b}) involving neutrons with higher energies, the MFPs for
both step-like and real cross-sections are nearly identical for these
neutrons, that results in the similarity between the patterns shown in
these figures.  Considering the proportions of both processes in the
production of cold neutrons, and accounting for the reduced
contributions of neutrons with longer wavelengths, we observe that for
the voluminous moderator, the single-collision process contributes to
around \qty{60}{\percent} of the cold neutron production, while for a
narrow high-aspect ratio moderator, this contribution is about
\qty{100}{\percent}~(\figref{fig:22:b}).
As a result, it is expected that the final
brightness gains for smaller moderator sizes will be approximately 1.6
times lower compared to the gains obtained from a model solely based
on the single-collision process. This is consistent with the
comparison of brightness gains between \figref{fig:16} and
\figref{fig:20}, where, for example, the maximal gains for
$H=\qty[round-precision=2]{0.01}{cm}$ are 17 and 11, respectively.

\section{Cold neutron spectrum emitted from low-dimensional para-hydrogen moderators}
\label{sec:5}

The change in relative contributions from single- and
multiple-collision processes, which depends on the moderator width as
described in the previous section, should clearly influence the
spectrum of cold neutrons emitted from such a moderator. Examples of
such spectra for a voluminous \TenByTenCm and a high-aspect ratio
\ZeroZeroOneByTenCm moderators are presented in \figref{fig:22} and
\figref{fig:23}, respectively, where the incident Maxwell
spectrum is also depicted. The corresponding curve is artificially
scaled to highlight the difference in shape between the thermal and
cold neutron spectra. The Maxwell spectrum used here is actually the
intensity spectrum calculated by scoring neutrons using the surface
crossing estimator~({\tt f1}).

Here it is important to note that the moderation process in para-hydrogen
differs fundamentally from the one observed in conventional neutron
moderators. In conventional moderators, high-energy neutrons are slowing
down by undergoing a series of successive collisions with the
moderator atoms, with each collision reducing the neutron's energy by
an average factor of $e$~(the slowing down process). At the final stage,
when neutron energy approaches the average kinetic energy of the
moderator atoms, neutrons undergo further multiple collisions
facilitating the exchange of kinetic energy between the neutrons and
the moderator atoms till the probability of down-scattering and
up-scattering become similar, so that neutrons become ``thermalized''.

In para-hydrogen, the main mechanism for neutron energy loss at the final
stage is not thermalization, but rather the slowing down because of
the transition between
rotational energy levels of the H$_2$ molecule, which are
separated by \qty{14.7}{\meV}~\cite{Wurz1973}~(this transition
is known as the para-to-ortho transition).

The fact that a neutron loses an entire \qty{14.7}{\meV} in a single
collision with para-hydrogen molecules makes the thermal-to-cold conversion
of neutrons with $E < \qty{30}{\milli\electronvolt}$ primarily a
single-collision process. Particularly, neutrons with energies in the
range of \qtyrange{15}{30}{\milli\electronvolt}, where the maximum of
the incident Maxwellian spectrum is located, essentially undergo the
single-collision process, because their final kinetic energy is less
than \qty{15}{\meV} and scattering cross-section for such neutrons is
very small. Hence, the spectrum of cold neutrons exiting the para-hydrogen
moderator is significantly depleted of neutrons with energies $E >
\qty{30}{\milli\electronvolt}$~($\lambda < \qty{1.65}{\angstrom}$), the
energy of these neutrons decreases to around \qty{15}{\meV}
($\lambda\approx\qty{2.34}{\angstrom}$), where the maximum of the
neutron spectrum is anticipated to be located. These considerations
are confirmed by the spectra shown in \figref{fig:22:a} and
\figref{fig:23:a}~(red curves), where the incident Maxwellian spectra
are depicted as black curves for illustration.

Thus, in low-dimensional para-hydrogen moderators of small width, where the
single-scattering process dominates, the Maxwellian spectrum observed
in the outgoing neutron beam does not stem from the thermalization
process typically observed in ortho-hydrogen or D$_2$ moderators. Instead, it
is a ``replica'' of the Maxwellian spectrum of the incident neutrons,
but with a shift towards lower energies.

\begin{figure}
\begin{subfigure}[T]{\figwidth}
\tikzsetnextfilename{main-figure39}
\begin{tikzpicture}
\end{tikzpicture}
\caption{}
\label{fig:22:a}
\end{subfigure} \hspace{3em}
\begin{subfigure}[T]{\figwidth}
\tikzsetnextfilename{main-figure40}
\begin{tikzpicture}
\end{tikzpicture}
\caption{}
\label{fig:22:b}
\end{subfigure}
\caption{(a)~Wavelength spectra of neutrons produced by single and
multiple collisions processes in a voluminous moderator of
\TenByTenCm. Black curve is for illustration purpose only, not
to scale. (b)~Fractions of neutrons corresponding to these processes
for each wavelength bin.}
\label{fig:22}

\todo[inline]{Use the same number of bins as in \figref{fig:18} and \ref{fig:19}}
\end{figure}
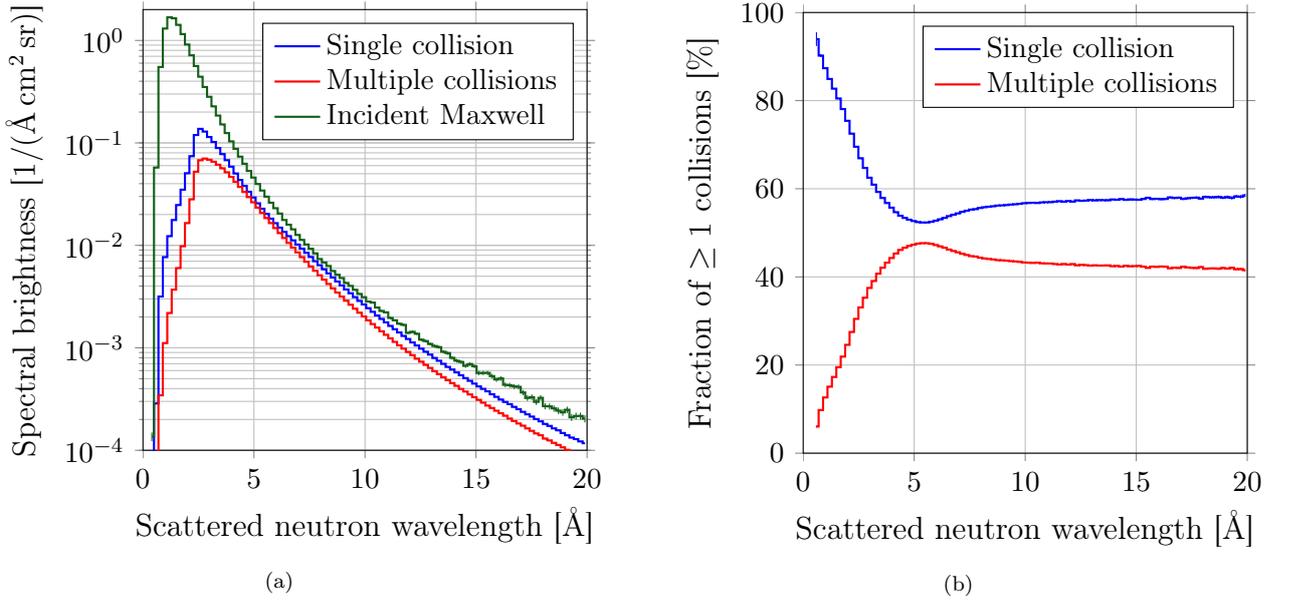
\begin{figure}
\centering
\begin{subfigure}[T]{\figwidth}
\tikzsetnextfilename{main-figure41}
\begin{tikzpicture}
\end{tikzpicture}
\caption{}
\label{fig:23:a}
\end{subfigure} \hspace{3em}
\begin{subfigure}[T]{\figwidth}
\tikzsetnextfilename{main-figure42}
\begin{tikzpicture}
\end{tikzpicture}
\caption{}
\label{fig:23:b}
\end{subfigure}
\caption{(a)~Wavelength spectra of neutrons produced by single and
multiple collisions processes in a high-aspect ratio moderator
of \ZeroZeroOneByTenCm. Black curve is for illustration purpose
only, not to scale. (b)~Fractions of neutrons corresponding to these processes
for each wavelength bin.}
\label{fig:23}
\todo[inline]{Either split into two separate figures or add a common caption}
\todo[inline]{Use the same number of bins as in \figref{fig:18} and \ref{fig:19}}
\todo[inline]{Improve statistics}
\todo{Rearrange label entries in \figref{fig:23:a}}
\end{figure}
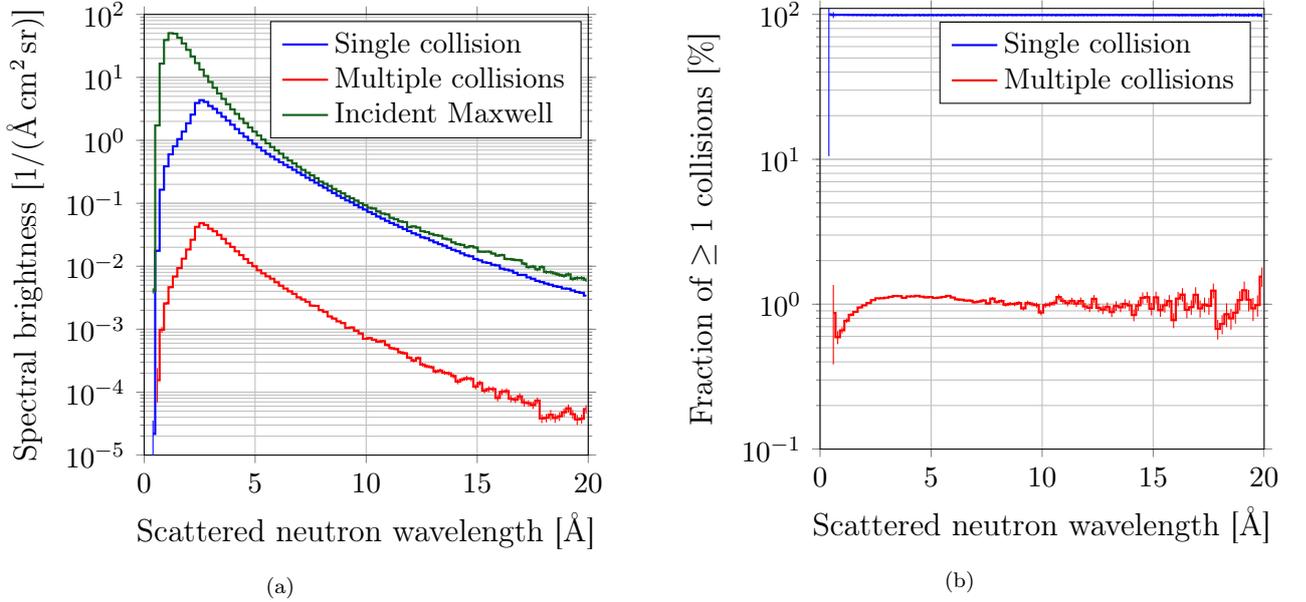

\section{Cold neutron source utilising assembly of low-dimensional moderators}
\label{sec:6}

\subsection{Neutron source concept}
\label{sec:6:1}

As discussed above, narrow high-aspect ratio moderators exhibit a
significant increase in brightness compared to wider voluminous
moderators. Moderators with enhanced brightness are particularly
advantageous for brightness-hungry instruments such as reflectometers
or small-angle neutron scattering~(SANS) diffractometers, where high
Q-resolution requires well-collimated neutron beams. In such cases,
the brightness gain directly contributes to the intensity at the
sample~\cite{Andersen2018}.

However, this enhanced brightness comes at the expense of reduced
intensity of emitted cold neutron beams due to the reduced moderator
volume. Indeed, these small yet bright moderators are not as
beneficial for intensity-hungry instruments with relaxed Q-resolution
requirements, e.g.\ like neutron spin echo~(NSE) or fundamental physics
instruments, as they produce fewer neutrons compared to conventional
voluminous moderators, albeit less bright.

In reference~\cite{Konik2023a}, an analysis was conducted to investigate how
moderator size affects the performance of neutron scattering
instruments. It was demonstrated that the best performance is achieved
with a moderator size not smaller than $D_{min}$, which is
instrument-specific. Using smaller moderators, though with
significantly enhanced brightness, leads nevertheless to the
degradation of instrument performance. Such moderators
under-illuminate the entrance of neutron guides that have a larger or
equal cross-section compared to the moderator size. This can lead to
irregularities in the structure of the outgoing beam divergence,
particularly observed in the case of reflectometer HERITAGE. To address this issue, a
combination of a cold moderator and a neutron guide of the same size
(\qty{3}{\cm}) has required the use of a multichannel neutron guide as a
partial solution~\cite{mattauch2017}. On the other hand, larger moderators ensure full
illumination of the neutron guide entrance, although they provide
lower beam brightness. Therefore, while larger moderators are
beneficial for intensity-hungry instruments, they may result in poorer
performance for brightness-hungry instruments.

Indeed, achieving both large moderator size and high brightness is
highly desirable. One possible approach is to increase the neutron
beam intensity by stacking multiple moderators adjacent to each
other. However, it is important to consider that, as demonstrated
earlier, the effective production of cold neutrons primarily occurs in
the vicinity of the moderator surfaces that are illuminated by thermal
neutrons. Since neighboring moderators can shadow each other, certain
surfaces may not be illuminated to the same extent as in the case of a
single moderator discussed earlier.

Let us consider two narrow moderators with brightnesses
$B_1=I_1/W$~(here we neglect the factor $\pi$~(solid angle) as it is
inconsequential to our current discussion). If these moderators are in
close contact~(\figref{fig:24:a}), the surfaces of the narrow moderators that
are in contact are not directly illuminated by thermal neutrons;
instead, they are shadowed by the adjacent moderator. Consequently,
the intensity $I'_1$ of the cold neutron beam produced in the shadowed
moderator is reduced compared to the intensity $I_1$ of the fully
illuminated single moderator. As a result, the average brightness
$B_{tot}$ of the moderator assembly also decreases:

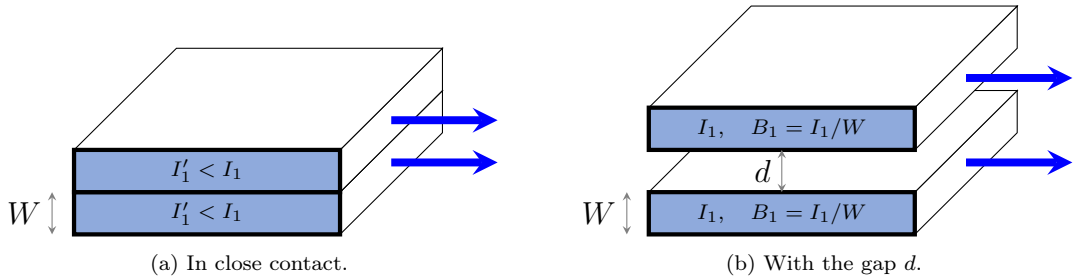
\begin{figure}[h]
\pgfmathsetmacro{\cubex}{5}
\pgfmathsetmacro{\cubey}{0.8}
\pgfmathsetmacro{\cubez}{5}
\pgfmathsetmacro{\d}{2}
\begin{subfigure}[t]{\figwidth}
\centering
\tikzsetnextfilename{main-figure43}
\begin{tikzpicture}
\end{tikzpicture}
\caption{In close contact.}
\label{fig:24:a}
\end{subfigure}
\begin{subfigure}[t]{\figwidth}
\centering
\tikzsetnextfilename{main-figure44}
\begin{tikzpicture}
\end{tikzpicture}
\caption{With the gap $d$.}
\label{fig:24:b}
\end{subfigure}
\caption{Arrangements of two 1-dimensional adjacent moderators.
Blue arrows depict the outgoing cold neutron beam.}
\label{fig:24}
\end{figure}

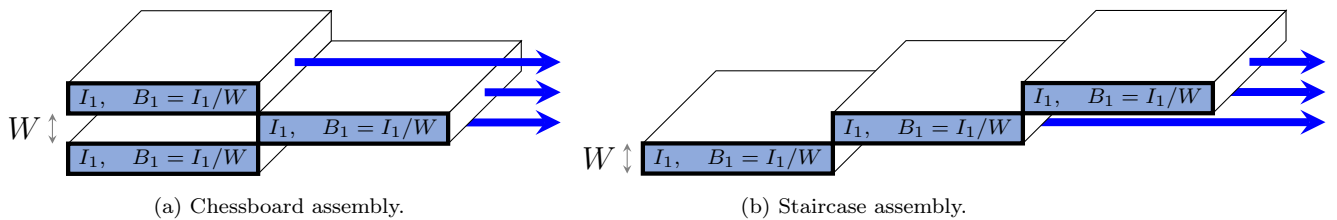
\begin{figure}[h]
\pgfmathsetmacro{\cubex}{5}
\pgfmathsetmacro{\cubey}{0.8}
\pgfmathsetmacro{\cubez}{5}
\pgfmathsetmacro{\d}{2}
\newcommand\ttt{\scriptsize{$I_1, \quad B_1=I_1/W$}\xspace}
\begin{subfigure}[t]{\figwidth}
\centering
\tikzsetnextfilename{main-figure45}
\begin{tikzpicture}
\end{tikzpicture}
\caption{Chessboard assembly.}
\label{fig:25:a}
\end{subfigure}
\begin{subfigure}[t]{\figwidth}
\centering
\tikzsetnextfilename{main-figure46}
\begin{tikzpicture}
\end{tikzpicture}
\caption{Staircase assembly.}
\label{fig:25:b}
\end{subfigure}
\caption{Assemblies of narrow moderators arranged like steps in
chessboard and staircase assemblies. Blue arrows depict the
outgoing cold neutron beam.}
\label{fig:25}
\end{figure}

\begin{equation}
B_{tot} \ne \dfrac{2I_1'}{2W} < B_1
\end{equation}

Actually, the stack of N moderators is equivalent to the moderator
with width $N\times W$ which brightness is reduced compared to that of
a single moderator~(as shown in \figref{fig:7}). For example, within a
stack of 10 narrow $(\num{0.3} \times \num{10})\,\unit{\cm\squared}$ moderators,
each could potentially provide a brightness of 8~(relative to the
brightness of a \TenByTenCm moderator) if considered
individually. However, the average brightness gain of the moderator
stack will be only around~2, equivalent to that of a single
\qty{3}{\cm} wide moderator~(see \figref{fig:7:b}).

If the moderators in the stack are separated by a gap~$d$, as shown
in~(\figref{fig:24:b}), the mutual shadowing of the moderator surfaces
is significantly reduced. When the gap~$d$ is approximately equal to
the moderator width~$W$, the mutual shadowing is reduced to a few
percent (for the typical moderator size of about \qty{10}{\cm}). In
this case, each moderator can be considered as practically fully
illuminated by thermal neutrons. However, from the perspective of a
neutron guide, this assembly of moderators appears as a moderator with
a dead area~(hole) of size $d$ at the center. As a result, the average
brightness of the moderator assembly is affected and is significantly
reduced compared to the brightness of a single moderator:

\begin{equation}
B_{tot} = \dfrac{2I_1}{2W + d} < B_1 = \dfrac{I_1}{W}
\end{equation}

Indeed, neither of  moderator assemblies shown in \figref{fig:24} provide any brightness
gain compared to the brightness of a single equally wide moderator.
To achieve a higher neutron intensity while maintaining the brightness
of a narrow moderator, it is crucial to ensure well-developed and
practically fully illuminated surfaces of the individual moderators in
the assembly, while avoiding any non-emitting areas or holes.

This can be achieved by filling the gap between adjacent moderators, which is shown in \figref{fig:24:b}, with an additional similar moderator, but shifted sideways by its length, thus creating a chessboard structure (\figref{fig:25:a}). One can also achieve the gap-free configuration by arranging moderators in a staircase structure (\figref{fig:25:b}). Such moderator assemblies has a large effective width $W_A$
equal to the total width of the closely stacked single
moderators~(e.g. $3W$ as shown in \figref{fig:25}). The losses in
thermal neutron illumination resulting from mutual shadowing are
rather small, less than \qty{10}{\percent}, and discussed in
\ref{app:B}.

By arranging $N$ single moderators like steps in the staircase or chessboard assembly,
the brightness of each individual moderator is effectively maintained,
while the intensities from each moderator are added up.

In a typical situation of the single cold para-hydrogen moderator, there is a
correlation between the brightness and size of moderators, with
specific brightness levels associated with particular moderator sizes,
and vice versa. However, when using an assembly of moderators, we
decouple these values: it becomes principally feasible for assemblies of any size
to achieve any desired brightness level, and conversely, any desired
brightness level can be achieved using assemblies of any size.

Suppose the brightness of a narrow moderator, $B_{narrow}$, is $k$
times higher than the brightness $B_{vol}$ of a voluminous one. Then,
a moderator assembly (i.e., cold neutron source) of $N$ individual
narrow moderators provides such increased cold neutron brightness
$B_{A}=B_{narrow}=kB_{vol}$, while having the same size as the
voluminous moderator $W_{vol} = W_A = NW$. Since intensity is the
product of brightness and size~(see \eqref{eq:Bcold}), the intensity
$I_A$ delivered by such a moderator assembly increases proportionally
to $N$ relative to intensity of individual narrow moderator: $I_A =
NI_{narrow}$. Thus, compared to the intensity of voluminous moderator
$I_{vol} = B_{vol}W_A$, the intensity $I_A$ increases by a factor of
$k$ as well:
\begin{equation}
I_A = NI_{narrow} = NB_{narrow}W = NkB_{vol}W_A/N = kI_{vol}.
\label{stair_equation}
\end{equation}
Indeed, the intensity delivered by the moderator assembly is increased
as much as its brightness with respect to voluminous moderator.
Detailed calculations of the intensity delivered by a moderator
assembly is presented in \secref{sec:6:2}.

Alternatively, a staircase geometry~(\figref{fig:25:b}) can be employed,
where the lateral shift of all individual moderators~(staircase steps) reduces
their shadowing by neighboring moderators. This geometry may prove to
be even more efficient, especially when used with thermal neutron
moderators of large length.

The concept of using a moderator assembly with increased size and
improved illumination addresses the need for enhancing the performance
of low Q-resolution intensity-hungry instruments at a low-dimensional
cold neutron source. This approach also effectively resolves the issue
of under-illumination at the entrance of a large neutron guide. For
example, in the case of the chessboard geometry shown in
\figref{fig:25:a}, the effective cold neutron source width would be $3W$~---
this ensures that the entrance of the neutron guide is sufficiently
illuminated, enabling a better quality of the outgoing neutron beam.

It is important to note that for pulsed neutron sources, the staircase
arrangement provides a notable increase in the effective neutron pulse
length, because the propagation time of cold neutrons along a
\qty{15}{\cm} long individual moderator is approximately
\qtyrange{0.1}{0.2}{\milli\second}. While this may not have a
significant impact on the pulse duration at long pulse neutron
sources, it is problematic for short pulse neutron source\footnote{The
authors are grateful to J.Voigt~(JCNS, Forschungszentrum Jülich) for
bringing this point to their attention.}.

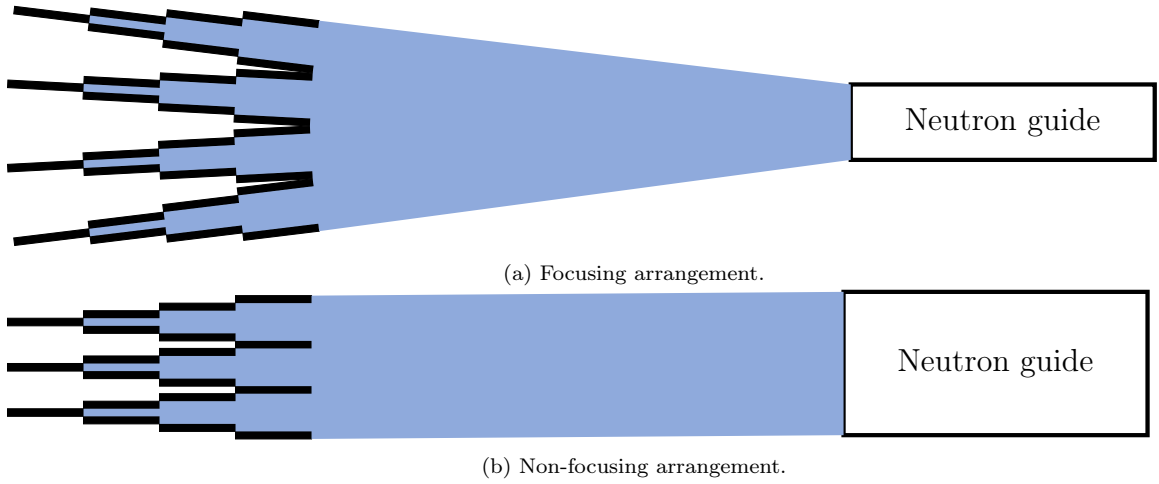
\begin{figure}
\centering
\pgfmathsetmacro\width{0.1}
\pgfmathsetmacro\length{1}
\pgfmathsetmacro\ngdist{7*\length} \pgfmathsetmacro\nglength{4*\length}
\pgfmathsetmacro\ngwidth{21*\width}
\begin{subfigure}{\linewidth}
\tikzsetnextfilename{main-figure47}
\begin{tikzpicture}
\end{tikzpicture}
\caption{Focusing arrangement.}
\end{subfigure}
\begin{subfigure}{\linewidth}
\tikzsetnextfilename{main-figure48}
\begin{tikzpicture}
\end{tikzpicture}
\caption{Non-focusing arrangement.}
\end{subfigure}
\caption{W-arrangements of single high-brightness moderators as a
cold neutron source.}
\label{fig:27}
\end{figure}

Certainly, narrow single high-brightness moderators can be assembled
in various configurations depending on the specific requirements and
design considerations. \Figref{fig:27} illustrates some possible
arrangements of these moderators, which can be used individually or
combined in more sophisticated designs such as $W$- or $V$-arrangements
for focusing or non-focusing purposes. These arrangements can be
useful in the design of a high-brightness cold neutron source,
particularly when aiming for a larger size and improved
performance. As indicated in the reference~\cite{Konik2023a}, there exists a
minimum cold neutron source size, denoted as $D_{min}$, for the optimal
performance of each particular instrument. Through the use of assembly
configurations, it becomes possible to employ several smaller
moderators with enhanced brightness to meet this $D_{min}$
requirement, resulting in an increased sample flux.

\subsection{Brightness and intensity of moderator assembly}

\begin{figure}
\centering
\hspace{-1cm}
\begin{subfigure}[t]{\figwidth}
\tikzsetnextfilename{main-figure49}
\begin{tikzpicture}
\end{tikzpicture}
\caption{}
\label{stair_brightness:a}
\end{subfigure}
\hspace{2cm}
\begin{subfigure}[t]{\figwidth}
\tikzsetnextfilename{main-figure50}
\begin{tikzpicture}
\end{tikzpicture}
\caption{}
\label{stair_brightness:b}
\end{subfigure}
\caption{Brightness and intensity gains of individual moderators
as a function of their width~$W$~(\subref{stair_brightness:a})
and number of steps~$N$~(\subref{stair_brightness:b}).  Orange
points represent intensity gains of a staircase or chessboard assembly with a
width $W_A$ of \qty{10}{\cm}. Solid lines are indicative.}
\label{stair_brightness}
\end{figure}
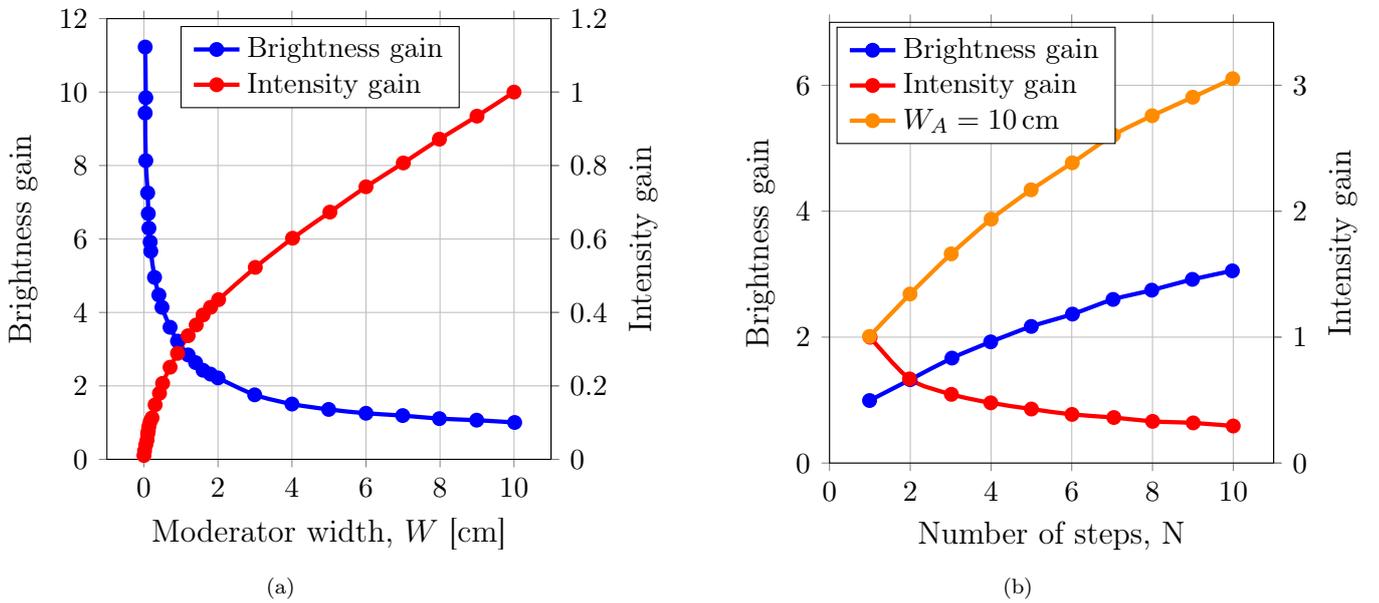

The intensity (total number of emitted neutrons) delivered by an
individual moderator is calculated as the product of brightness and
the moderator width. As depicted in Fig.~\ref{stair_brightness:a}, a
moderator with a width $W < \qty{2}{\cm}$ provides a substantial \numrange{2}{3} times
increase in brightness compared to a voluminous moderator with
dimensions of \TenByTenCm (black squares) by the price of
the delivered intensity being reduced by a factor of approximately 3
due to the smaller size of the moderator~(Fig.~\ref{stair_brightness:a}, red circles).

However, for an assembly of moderators with a total width of $W_A$,
this reduction is completely compensated by the number $N = W_A/W$ of
moderators in an assembly. This is illustrated by
Fig.~\ref{stair_brightness:b}, where brightness and intensity gains
are plotted as a function of the number of steps $N \in [1-10]$ in a
$W_A=\qty{10}{\cm}$ assembly.  Here black squares and red circles
correspond to individual moderator (as in
Fig.~\ref{stair_brightness:a}), while blue triangles depict the
intensity gains of the $N$-step staircase assembly and all values are
normalized to the respective values of the voluminous moderator with
the same width $W_A=\qty{10}{\cm}$.

It is important to note, that brightness and intensity gains in
Fig.~\ref{stair_brightness:b} are equal, as it was already shown
previously in~\secref{sec:6:1} (see Eq.~(\ref{stair_equation}) and
corresponding discussion). Indeed, the staircase and chessboard
moderator assemblies provide a simultaneous, significant, and equal
increase in both brightness and intensity, which is a distinct feature
of this type of cold neutron source.

\subsubsection{Procedure for brightness gain calculations}

In the previous subsection, we presented the concept of chessboard and
staircase moderator assemblies, each consisting of several individual
moderators~(steps). The overall brightness gain $G_{assembly}$ for
such an assembly, compared to a voluminous moderator, can be
calculated as the product of five independent factors:

\begin{equation}
G_{assembly} = G_{compression} \cdot G_{length} \cdot F_{position} \cdot F_{wall} \cdot F_{shadow},
\label{gain_equation}
\end{equation}
where $G_{compression}$ represents the compression gain of the
moderator compared to a voluminous moderator with similar outer
dimensions $W_A$ and $H$, $G_{length}$ is the length gain of the
elongated para-hydrogen moderator, $F_{position}$ accounts for the
non-uniformity in the incoming thermal neutron flux, $F_{wall}$
accounts for wall losses and $F_{shadow}$ accounts for the mutual
shadowing of individual moderators. Below, we will discuss the first
four factors, leaving the topic of shadowing for more detailed
consideration in \ref{app:B}.

\begin{itemize}
\item Moderator compression.

Fig.~\ref{fig:20} shows MCNP calculated brightness gains
$G_{compression}$ depending on the moderator width. All results are
normalized for \TenByTenCm moderator. Results corresponding to
moderator height \qty{10}{\cm} were fitted and that function was used for
further calculations.

It's important to note, that due to identical outer dimensions of the
entire moderator assembly, it can be placed in the same beam tube as a
wide individual moderator, and does not require the modification of
the position of surrounding reflectors. Indeed, the dependence of
brightness gain on moderator sizes for individual moderators in
assemblies is the same as for a voluminous moderator (see
\figref{fig:20}).

\item Moderator length.

Para-hydrogen exhibits a very low scattering cross-section for cold
neutrons. MFP for neutrons with
$\num{12}<E<\qty{25}{\milli\electronvolt}$, where maximum of cold
neutron spectrum is situated, can be estimated as $\lambda_{cold} =
\qty{10}{\centi\meter}$ on average~(\figref{fig:8}). Indeed, the
moderator brightness gain due to increased length can be estimated as

\begin{equation}
G_{length} = \int_0^L b(l) \, e^{-\lambda_{cold}/l} dl,
\end{equation}
where $b(l)$ is the thermal neutron brightness in the position~$l$.

\item Thermal neutron flux distribution.

As mentioned earlier, the thermal neutron flux may vary along the
moderator assembly. This variation imposes a limitation for the
proposed moderator geometries, as the reduced thermal neutron
illumination at the periphery of the staircase moderator assembly
limits its feasible overall length and the number of individual
moderators that can be used in the assembly~(\figref{fig:thermal}).

This limitation can be partially overcome by using multiple parallel
staircase and chessboard assemblies, as shown in
Figures~\ref{multi_stair_config:c} and~\ref{multi_stair_config:d},
respectively. In these figures, the overall assembly width $W_A$ is
kept constant, while the step width in multiple parallel staircases
configuration is reduced. This leads to an increase in brightness from
each step, although within limit defined by the wall thickness, as
discussed below.

\begin{figure}
\def\width{12mm}
\def\height{0.33*\width}
\centering
\tikzsetnextfilename{main-figure51}
\begin{tikzpicture}
\end{tikzpicture}
\caption{Thermal flux distribution for a typical thermal heavy water
research reactor~(shown in black) superimposed with the staircase
moderator assembly layout.}
\label{fig:thermal}
\end{figure}
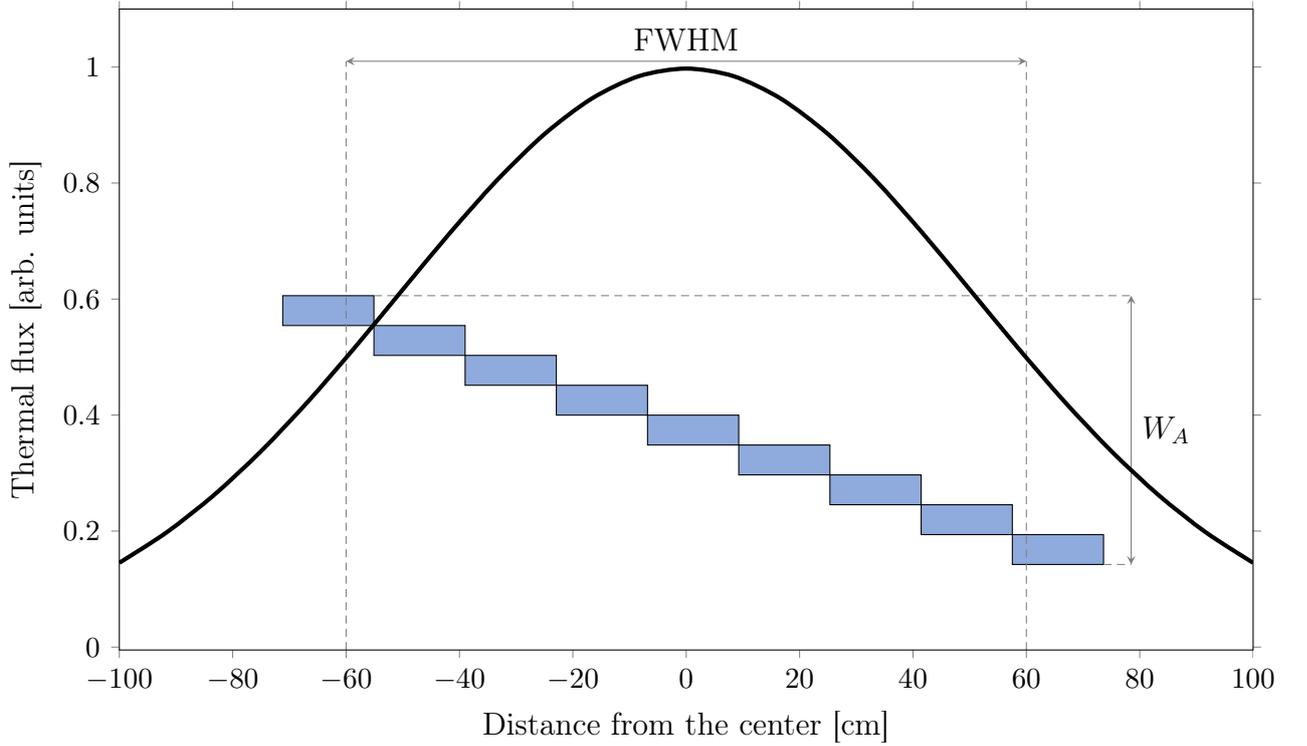

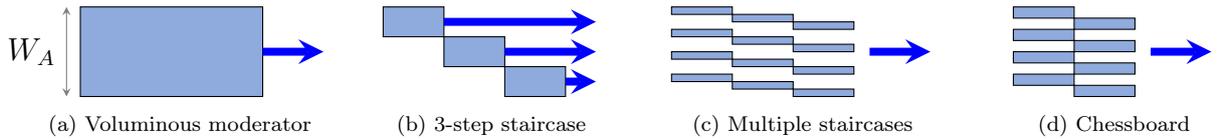
\begin{figure}[t]
\centering
\def\width{8mm}
\def\height{0.495*\width}
\begin{subfigure}{0.24\linewidth}
\hspace*{-7mm}
\tikzsetnextfilename{main-figure52}
\begin{tikzpicture}
\end{tikzpicture}
\caption{Voluminous moderator}
\label{multi_stair_config:a}
\end{subfigure}
\begin{subfigure}{0.24\linewidth}
\centering
\tikzsetnextfilename{main-figure53}
\begin{tikzpicture}
\end{tikzpicture}
\caption{3-step staircase}
\label{multi_stair_config:b}
\end{subfigure}
\begin{subfigure}{0.24\linewidth}
\centering
\tikzsetnextfilename{main-figure54}
\begin{tikzpicture}
\end{tikzpicture}
\caption{Multiple staircases}
\label{multi_stair_config:c}
\end{subfigure}
\begin{subfigure}{0.24\linewidth}
\centering
\tikzsetnextfilename{main-figure55}
\begin{tikzpicture}
\end{tikzpicture}
\caption{Chessboard}
\label{multi_stair_config:d}
\end{subfigure}
\caption{Layouts of various moderator assemblies of the same total
width $W_A$. Blue arrows indicate outgoing cold neutron beam
direction.}
\label{multi_stair_config}
\end{figure}

To incorporate the corresponding factor $F_{position}$ into
calculations, the thermal flux distribution along the beam tube must
first be defined. In the following sections, we will consider two
types of distributions: one for heavy water reactor sources, with a
typical full width at half maximum~(FWHM) of \qty{120}{\cm}~(Figures
\ref{fig:thermal} and \ref{fig:reactor:layouts}) and for
spallation/compact neutron sources, with a typical FWHM of
\qty{25}{\cm}~(\figref{fig:spallation:layouts}).

\item Wall losses.

One of the factors decreasing the potentially achievable assembly
brightness is the finite wall thickness of each individual moderator,
which causes a reduction in the effective emitting area of the
assembled cold neutron source. The losses can be minimised when the
horizontal walls of neighbouring moderators overlap with each other, as
depicted in Fig.~\ref{fig:28:a}. Losses due to walls are proportional
to the percentage of useful bright surface covered by walls:

\begin{equation}
F_{wall} = 1 - a \, \dfrac{N-1}{W_A},
\end{equation}
where $F_{wall}$ is the remaining useful relative brightness, $a$ is
the wall thickness, $N$ is the number of steps and $W_A$ is the
overall moderator width.

By multiplying this factor with the brightness gain dependence on the
single moderator width $W$~(as shown in \figref{fig:18}), we can
determine the reduced brightness gain for different wall thicknesses,
which is presented in \figref{fig:28:b} for the moderator height $H =
\qty{10}{\cm}$. While the brightness gain for a single moderator
increases monotonically as its size decreases, considering the changes
in wall thickness alters this dependence. For a given wall thickness,
there exists a maximum achievable brightness gain, and this value
increases as the walls become thinner. The step width should not be
less than \qtyrange[round-precision=2]{0.25}{0.5}{\cm}~(depending on wall thickness),
limiting the allowable total number of steps in the assembly, e.g., to
12 for $W_A=\qty{3}{\cm}$ and 40 for $W_A=\qty{10}{\cm}$~(for a wall
thickness of \qty{1}{\mm}).

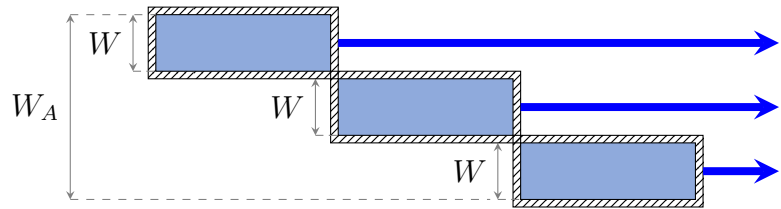
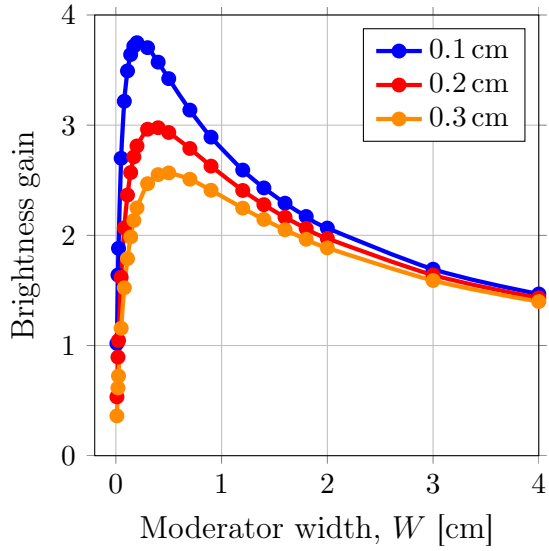
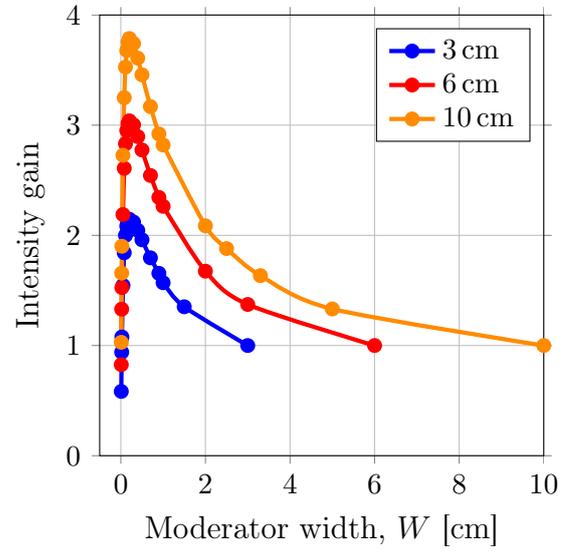
\begin{figure}
\centering
\begin{subfigure}[t]{\figwidth}
\centering
\tikzsetnextfilename{main-figure56}
\begin{tikzpicture}
\end{tikzpicture}
\caption{Overlap of horizontal walls of neighboring moderators.}
\label{fig:28:a}
\end{subfigure}

\begin{subfigure}[t]{\figwidth}
\tikzsetnextfilename{main-figure57}
\begin{tikzpicture}
\end{tikzpicture}
\caption{Brightness gain for various moderator wall thicknesses,
normalised to the \TenByTenCm moderator brightness (calculated
based on the Fig.~\ref{fig:20} data). Solid lines are indicative.}
\label{fig:28:b}
\end{subfigure}
\hspace{2em}
\begin{subfigure}[t]{\figwidth}
\tikzsetnextfilename{main-figure58}
\begin{tikzpicture}
\end{tikzpicture}
\caption{Intensity gains achieved by moderator assemblies~(with a
wall thickness of~\qty{1}{\mm}) compared to the intensity of
voluminous moderators with the same width
$W_A$~(\qtylist[list-units=single]{3;6;10}{\cm}). Solid lines are indicative.}
\label{fig:28:c}
\end{subfigure}
\caption{Wall thickness effect.}
\label{fig:28}
\end{figure}

Since the intensity gain of such assembly is calculated as the product
of brightness and assembly width $W_A$, the finite wall thickness
results in non-monotonic behavior of the intensity gain. It is
illustrated in \figref{fig:28:c}, where intensity gains for different
$W_A$ are shown. One can see, that intensity gain for a \qty{10}{\cm} wide
assembly consisting of 6 single \qty{1.6}{\cm} wide moderators is
approximately 2.5 relative to the \qty{10}{\cm} wide voluminous moderator, and
for a \qty{6}{\cm} wide assembly consisting of 4 single \qty{1.5}{\cm} wide
moderators, the intensity gain is about 2 relative to the \qty{6}{\cm} wide
voluminous moderator.

Consequently, using thinner walls proves to be advantageous. However,
as para-hydrogen in the moderator should be under high pressure, up to
\qty{20}{\bar}~\cite{Zanini2019a}, the thickness of the moderator walls should
be sufficient for the safe operation of such moderators. Certainly,
minimising the wall thickness poses a significant challenge and
engineering efforts are needed to address it. One potential approach
is to manufacture the single moderator from a single aluminium piece
with internal ribs as shown in Fig.~\ref{fig:30}. These ribs serve the
dual purpose of facilitating the proper flow of liquid hydrogen and
acting as connectors between the large area surfaces of the moderator
(approximately \numproduct{10x15}\,\unit{\cm\squared}). This design enables the
desired flow characteristics while potentially reducing the wall
thickness to around \qty{1}{\mm}, thereby minimising the impact on brightness
gain.

\begin{figure}[h!]
\begin{subfigure}{\figwidth}
\includegraphics[width=\linewidth]{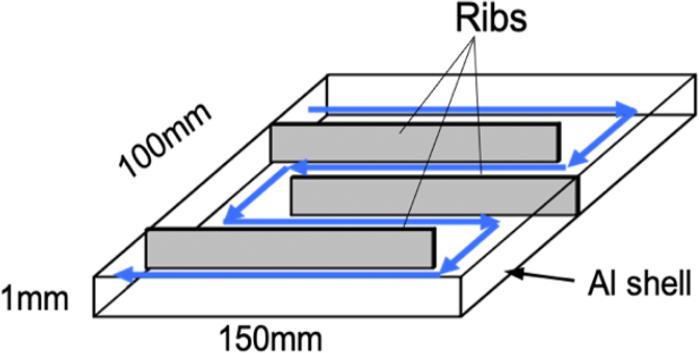}
\vspace{0.75cm}
\caption{3D view}
\label{fig:30:a}
\end{subfigure}
\hspace{3em}
\begin{subfigure}{\figwidth}
\includegraphics[width=\linewidth]{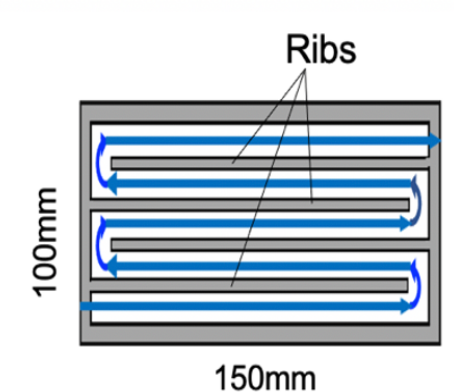}
\caption{Top view}
\label{fig:30:b}
\end{subfigure}
\caption{Conceptual drawing of thin narrow elongated cold moderator
made of a single aluminum piece with vertical ribs connecting the
thin upper and down large area surfaces. This design was suggested
by Y.\,Be{\ss}ler~(Central Institute of Engineering, Electronics and
Analytics~(ZEA), Forschungszentrum Jülich).}
\label{fig:30}
\end{figure}

\end{itemize}

\subsection{Results}
\label{sec:6:2}

\subsubsection{Moderator assemblies at reactor source}

First, consider the brightness gains of single staircase moderator
assemblies with different numbers of steps. These assemblies are
installed in the beam tube of a heavy water research reactor with a
thermal flux distribution of FWHM = \qty{120}{\cm}~(as shown in
\figref{fig:reactor:layouts}). The results of calculations are presented in
Fig.~\ref{overall1}. All calculations are performed using
Eq.~(\ref{gain_equation}), with the outer dimensions of the entire
moderator assembly remaining constant.

\begin{figure}
\def\width{12mm}
\def\height{0.33*\width}
\centering
\tikzsetnextfilename{main-figure59}
\begin{tikzpicture}
\end{tikzpicture}
\caption{Typical thermal neutron flux distributions in a heavy water
reactor~(black curve), along with the placement of various
moderator assemblies of equal overall width $W_A$.}
\label{fig:reactor:layouts}
\end{figure}
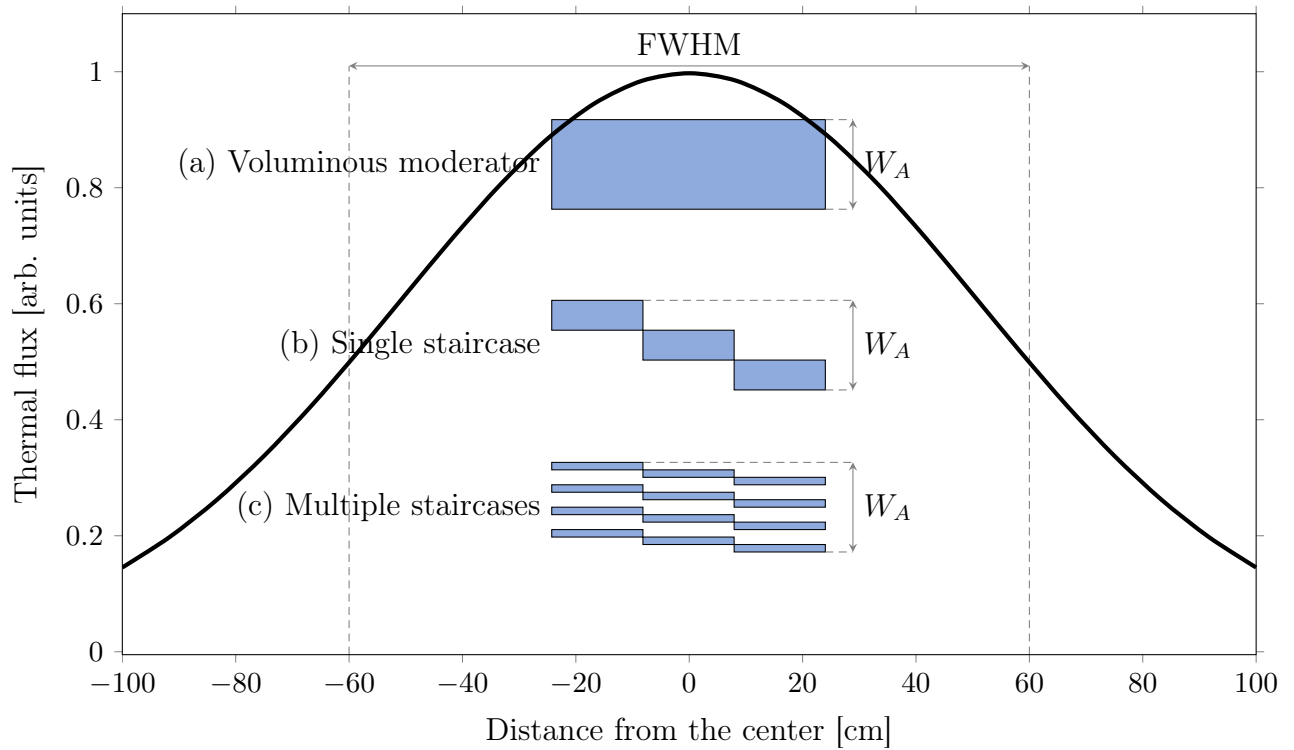

\begin{figure}
\def\width{7mm}
\def\height{0.1*\width}
\centering
\begin{subfigure}{\figwidth}
\tikzsetnextfilename{main-figure60}
\begin{tikzpicture}
\end{tikzpicture}
\caption{$W_A = \qty{3}{\cm}$}
\label{overall1:3cm}
\end{subfigure}
\begin{subfigure}{\figwidth}
\def\newheight{3.333*\height}
\tikzsetnextfilename{main-figure61}
\begin{tikzpicture}
\end{tikzpicture}
\caption{$W_A = \qty{10}{\cm}$}
\label{overall1:10cm}
\end{subfigure}
\caption{Overall brightness and intensity gain of various single staircase moderator
assemblies with \qty{1}{\mm} wall thickness installed in a heavy
water reactor as a function of step length.}
\label{overall1}
\end{figure}
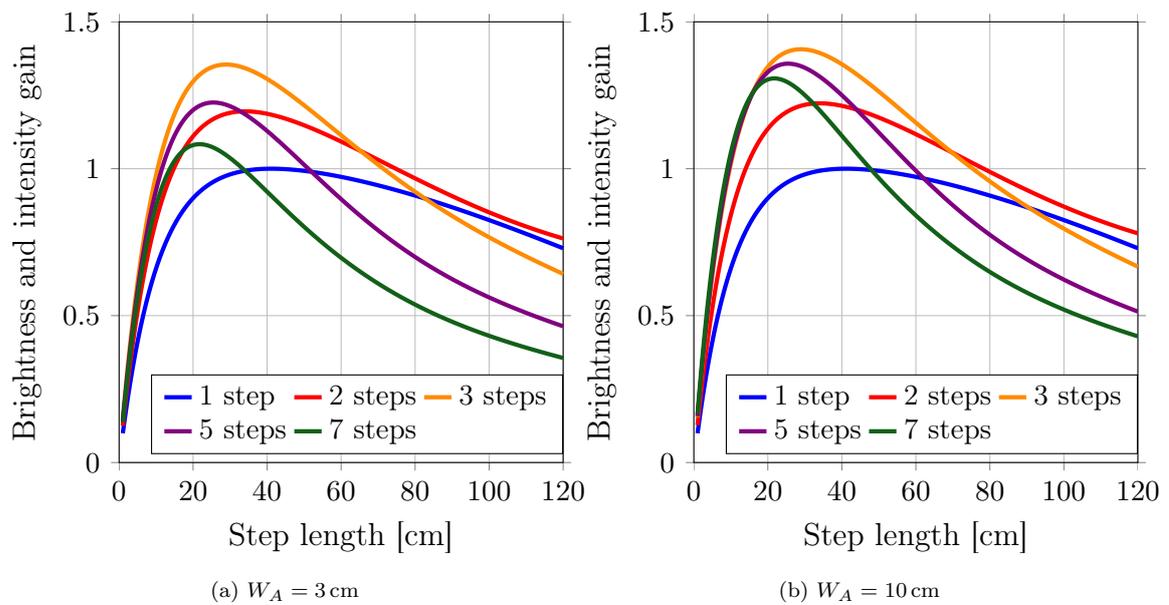

The blue lines in Fig.~\ref{overall1} correspond to a 1 step
moderator, essentially representing the behavior of a voluminous
moderator~(depicted as (a) in Fig.~\ref{fig:reactor:layouts}) with $H=\qty{10}{\cm}$
and varying length. Its brightness increases with increased length
until a certain point, after which it slowly decreases. This decrease
is due to the following reason: since cold neutrons can only be
emitted from the moderator's depth not exceeding 3 MFPs for cold
neutrons, the emitting part of a long moderator is located outside the
maximum thermal flux distribution region. The maximum of this 1 step
function is taken as 1 and all other brightness functions shown in
Fig.~\ref{overall1} are normalized by this value.

Fig.~\ref{overall1} also shows the overall brightness gains for single
staircase moderator assemblies (depicted as (b) in
Fig.~\ref{fig:reactor:layouts}) with a width $W_A$ of \qty{3}{\cm} or \qty{10}{\cm} and a
wall thickness of \qty{1}{\mm}. The calculated brightness of the assemblies is
the average brightness over all steps. One can see that calculations
are supporting the physical considerations in Sect. \ref{sec:6:1} and in general
single staircase assemblies provide higher gains compared to the
voluminous moderator. The 3-step configuration is most preferable,
providing a relative gain of about 1.35 for \qty{3}{\cm} and 1.4 for \qty{10}{\cm}
wide assemblies, so that their width doesn't significantly influence
the results. Brightness gains for smaller $W_A$ are slightly lower
because moderators' walls occupy a larger relative part of the total
assembly width.

\begin{figure}
\def\width{7mm}
\def\height{0.1*\width}
\centering
\begin{subfigure}{\figwidth}
\tikzsetnextfilename{main-figure62}
\begin{tikzpicture}
\end{tikzpicture}
\caption{$W_A = \qty{3}{\cm}$}
\label{overall2:3cm}
\end{subfigure}
\begin{subfigure}{\figwidth}
\def\newheight{3.333*\height}
\tikzsetnextfilename{main-figure63}
\begin{tikzpicture}
\end{tikzpicture}
\caption{$W_A = \qty{10}{\cm}$}
\label{overall2:10cm}
\end{subfigure}
\caption{Overall brightness and intensity gain of various 4-staircase moderator
assemblies with \qty{1}{\mm} wall thickness installed in a heavy
water reactor as a function of step length. }
\label{overall2}
\end{figure}
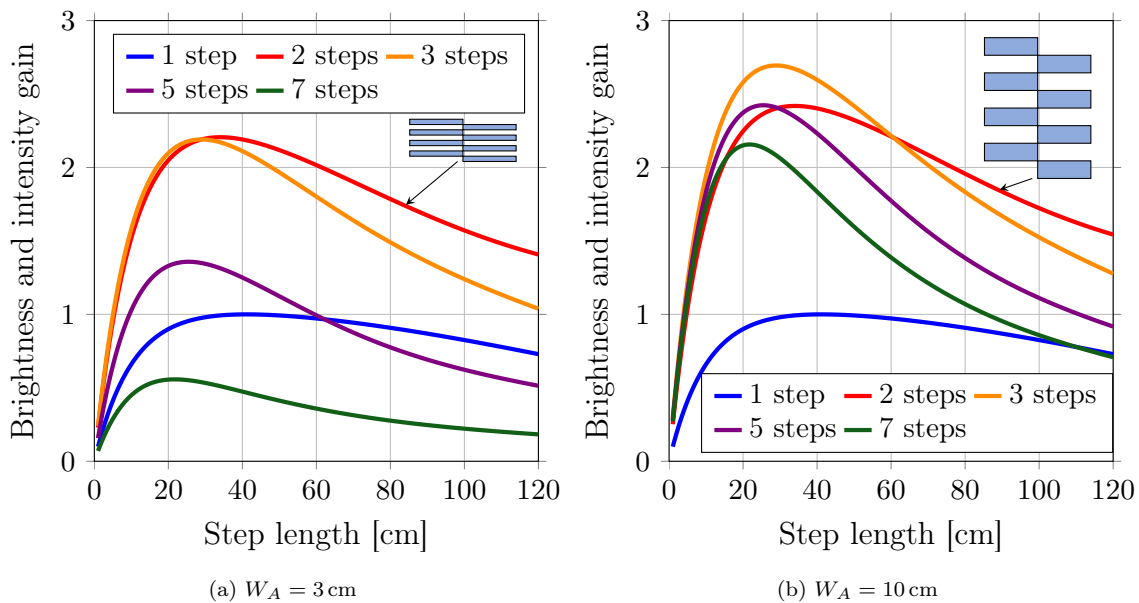

Multiple parallel staircases assemblies provide higher gains than a
single staircase arrangement. Fig.~\ref{overall2} shows potential
gains for 4-staircase assemblies with an overall moderator width of
$W_A=\qty{3}{\cm}$~(\subref{overall2:3cm}) and
$W_A=\qty{10}{\cm}$~(\subref{overall2:10cm}). As before, the blue
lines in these figures correspond to individual moderators of the same
width and are used as a reference. As a visual aid, the two-step
configurations corresponding to the red curves are depicted.

One can see that substantial additional gains can be achieved, e.g.,
$G_{assembly}$ up to 2.7 for $W_A=\qty{10}{\cm}$ moderator assembly with 3
steps in each of 4 staircases (an additional factor of about 2
compared to a single staircase configuration). Such an assembly would
consist of 12 individual moderators, each with a width of \qty{0.8}{\cm}, so
that they provide high brightness.

For $W_A=\qty{3}{\cm}$, the gains are approximately 2.2 for both
2-step and 3-step staircases, each having different step lengths
(note, that 2-step $N$-staircase configuration is essentially a
chessboard configuration with $2N$ number of steps). The total length
of the optimal assembly is about \qty{70}{\cm}. It's noteworthy that,
for the $W_A=\qty{3}{\cm}$ moderator assembly with 5 steps on each of
4 stairs, the step width is only
\qty[round-precision=2]{0.15}{\cm}. This corresponds to a
significantly reduced brightness (see Fig.~\ref{fig:28:b} and
\subref{fig:28:c}) due to the wall effect, resulting in a substantial
decrease in the brightness of such an assembly.

Notably, even relatively simple moderator assemblies composed of 4
individual moderators in chessboard geometry yield significant
enhancements in brightness and intensity of more than \qty{70}{\percent}, or gain of
almost 1.7~(red curve in Fig.~\ref{overall3}).

\begin{figure}
\def\width{7mm}
\def\height{0.333*\width}
\centering
\tikzsetnextfilename{main-figure64}
\begin{tikzpicture}
\end{tikzpicture}
\caption{Overall brightness and intensity gain of various 2-staircase moderator
assemblies with $W_A=\qty{10}{\cm}$ and \qty{1}{\mm} wall thickness installed in a heavy
water reactor as a function of step length.}

\label{overall3}
\end{figure}

\subsubsection{Moderator assemblies at spallation or compact sources}

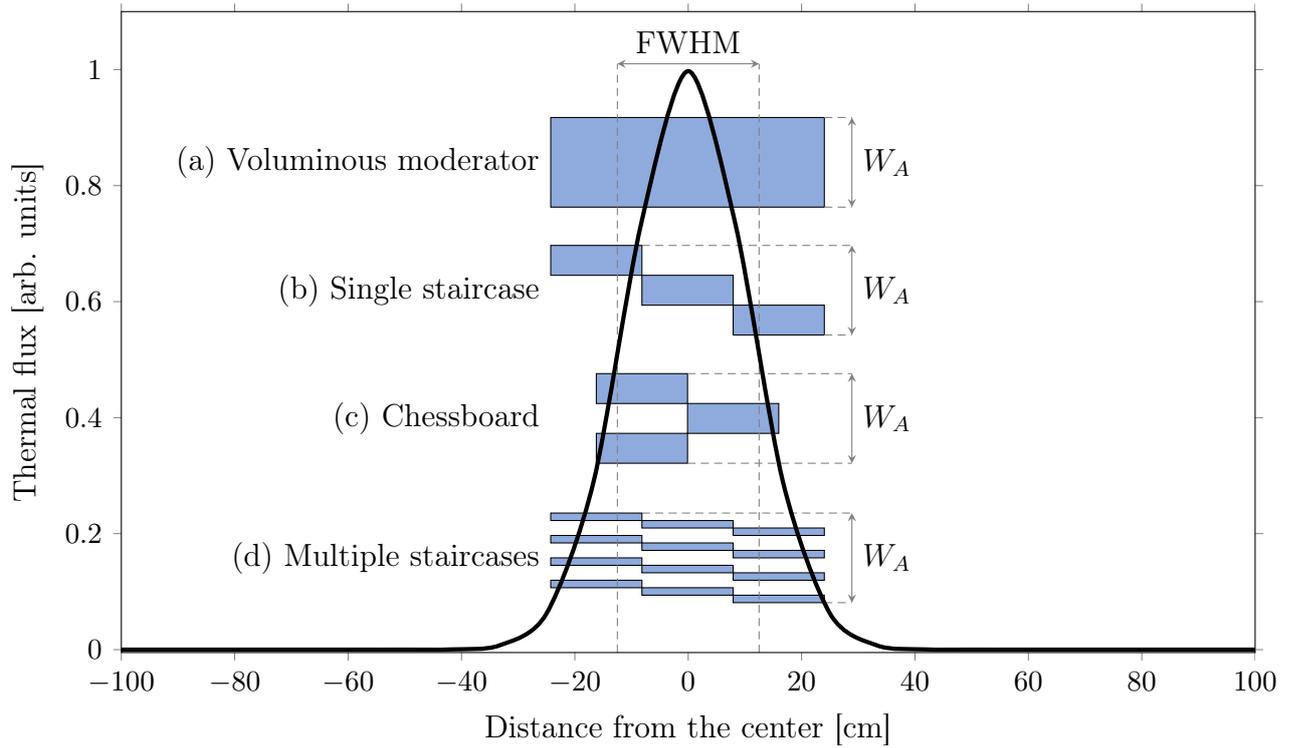
\begin{figure}
\def\width{12mm}
\def\height{0.33*\width}
\centering
\tikzsetnextfilename{main-figure65}
\begin{tikzpicture}
\end{tikzpicture}
\caption{Typical thermal neutron flux distributions in spallation
and compact neutron sources~(black curve), along with the placement of various moderator
assemblies of equal overall width $W_A$.}
\label{fig:spallation:layouts}
\end{figure}

Now we examine a more compact thermal flux distribution, typical for
spallation and compact neutron sources, and consider the performance of
moderator assemblies illuminated by a thermal flux distribution with
FWHM = \qty{25}{\cm}, as shown by the black curve in
Fig.~\ref{fig:spallation:layouts}. As before, all calculations are
performed using Eq.~(\ref{gain_equation}), with the outer dimensions
of the entire moderator assembly remaining constant.

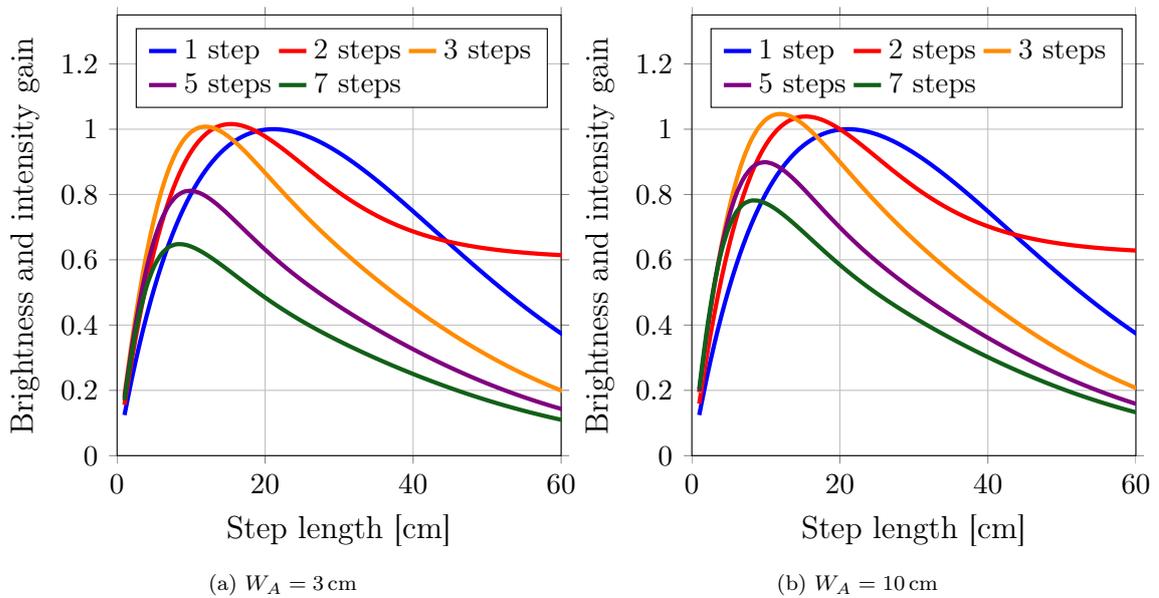
\begin{figure}
\def\width{7mm}
\def\height{0.1*\width}
\centering
\begin{subfigure}{\figwidth}
\tikzsetnextfilename{main-figure66}
\begin{tikzpicture}
\end{tikzpicture}
\caption{$W_A = \qty{3}{\cm}$}
\label{overall4:3cm}
\end{subfigure}
\begin{subfigure}{\figwidth}
\def\newheight{3.333*\height}
\tikzsetnextfilename{main-figure67}
\begin{tikzpicture}
\end{tikzpicture}
\caption{$W_A = \qty{10}{\cm}$}
\label{overall4:10cm}
\end{subfigure}
\caption{Overall brightness and intensity gain of various single staircase moderator
assemblies with \qty{1}{\mm} wall thickness installed in a spallation or compact neutron source as a function of step length.}

\label{overall4}
\end{figure}

Fig.~\ref{overall4} shows brightness gains for single staircase
moderator assemblies with an overall width of $W_A=\qty{3}{\cm}$ and
$W_A=\qty{10}{\cm}$, each with a wall thickness of \qty{1}{\mm}. Similar to
plots in Fig.~\ref{overall1}, the blue lines~(1 step) represent
the behavior of a voluminous moderator, which is depicted as (a) in
Fig.~\ref{fig:spallation:layouts}. Note the sharp decrease in brightness~(in
contrast to the reactor case considered above) after a certain
length. All brightness functions shown in \figref{overall4} are
normalised by the maximum value of the 1 step function for the
corresponding assembly width. One can observe a notable reduction in
achievable gains for single staircase configurations compared to the
reactor case (Fig.~\ref{overall1}), as there is limited space to shift
individual moderators along the cold neutron beam axis. In this case a
single staircase moderator assembly demonstrates no gain over
conventional moderators.

A substantial enhancement in brightness can be achieved by employing
moderator assemblies configured in chessboard or multiple parallel
staircases geometry~(see Fig.~\ref{multi_stair_config}). In
Figs.~\ref{overall5} and \ref{overall6}, brightness gains for
chessboard and 4-staircase moderator assemblies (depicted in
Fig.~\ref{fig:spallation:layouts} as~(c) and~(d), respectively) with overall
widths of $W_A=\qty{3}{\cm}$ and $W_A=\qty{10}{\cm}$ are presented. Similar to
previous plots, these results are normalised by the maximum value of
the 1 step function for the corresponding assembly width.  One can
see that for a simple V-shaped, 3-step chessboard moderator
assembly, gains of about 1.4 are achievable. Note that since a
V-shaped assembly consisting of three steps is not symmetrical with
respect to the vertical axis, the gains for the V pointing in the
outgoing beam direction~(e.g. towards a neutron guide) are higher
than in the opposite direction. This is because, in the first case,
the end parts of the two moderators placed in the maximum thermal
neutron flux distribution emit cold neutrons without further
propagation through a thick layer of para-hydrogen, whereas in the
second case, this occurs only for one moderator.

For the 4-staircase moderator assemblies~(Fig.~\ref{overall5}) gains
about \numrange{1.9}{2.1} are achievable. The staircases in optimal
moderator assemblies would consist of 2 steps with step length
$l_{step}\approx \qty{18}{\cm}$, each with a width of \qty[round-precision=2]{0.37}{\cm}
for $W_A=\qty{3}{\cm}$ and \qty{1.2}{\cm} for $W_A=\qty{10}{\cm}$, so
that they provide high brightness~(see Fig.~\ref{fig:20}).
Note, that for $W_A=\qty{3}{\cm}$, 5-step staircases will have the
step width of \qty[round-precision=2]{0.15}{\cm}, which corresponds to a significantly
reduced brightness of individual moderators with the wall thickness of
\qty{0.1}{\cm} and, consequently, in the absence of gain.

\begin{figure}
\def\width{7mm}
\def\height{0.1*\width}
\centering
\begin{subfigure}{\figwidth}
\tikzsetnextfilename{main-figure68}
\begin{tikzpicture}
\end{tikzpicture}
\caption{$W_A = \qty{3}{\cm}$}
\label{overall6:3cm}
\end{subfigure}
\begin{subfigure}{\figwidth}
\def\newheight{3.333*\height}
\tikzsetnextfilename{main-figure69}
\begin{tikzpicture}
\end{tikzpicture}
\caption{$W_A = \qty{10}{\cm}$}
\label{overall6:10cm}
\end{subfigure}
\caption{Overall brightness and intensity gain of chessboard moderator
assemblies with \qty{1}{\mm} wall thickness installed in a spallation
or compact neutron source as a function of step length.}
\label{overall6}
\end{figure}
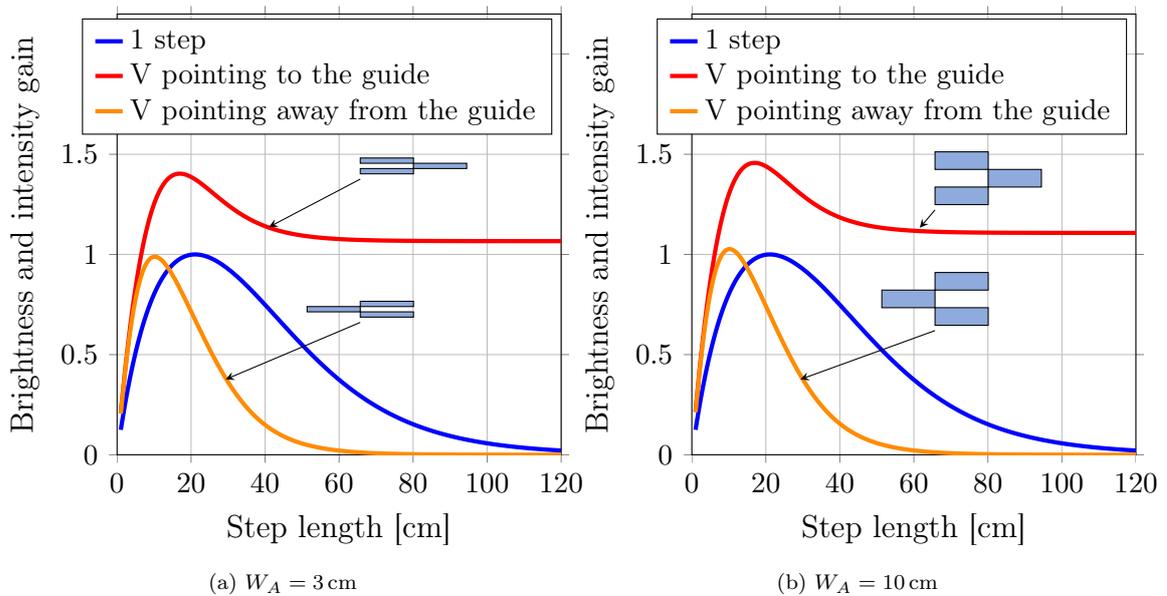

\begin{figure}
\def\width{7mm}
\def\height{0.1*\width}
\centering
\begin{subfigure}{\figwidth}
\tikzsetnextfilename{main-figure70}
\begin{tikzpicture}
\end{tikzpicture}
\caption{$W_A = \qty{3}{\cm}$}
\label{overall5:3cm}
\end{subfigure}
\begin{subfigure}{\figwidth}
\def\newheight{3.333*\height}
\tikzsetnextfilename{main-figure71}
\begin{tikzpicture}
\end{tikzpicture}
\caption{$W_A = \qty{10}{\cm}$}
\label{overall5:10cm}
\end{subfigure}
\caption{Overall brightness and intensity gain of 4-staircase moderator
assemblies with \qty{1}{\mm} wall thickness installed in a spallation
or compact neutron source as a function of step length.}

\label{overall5}
\end{figure}
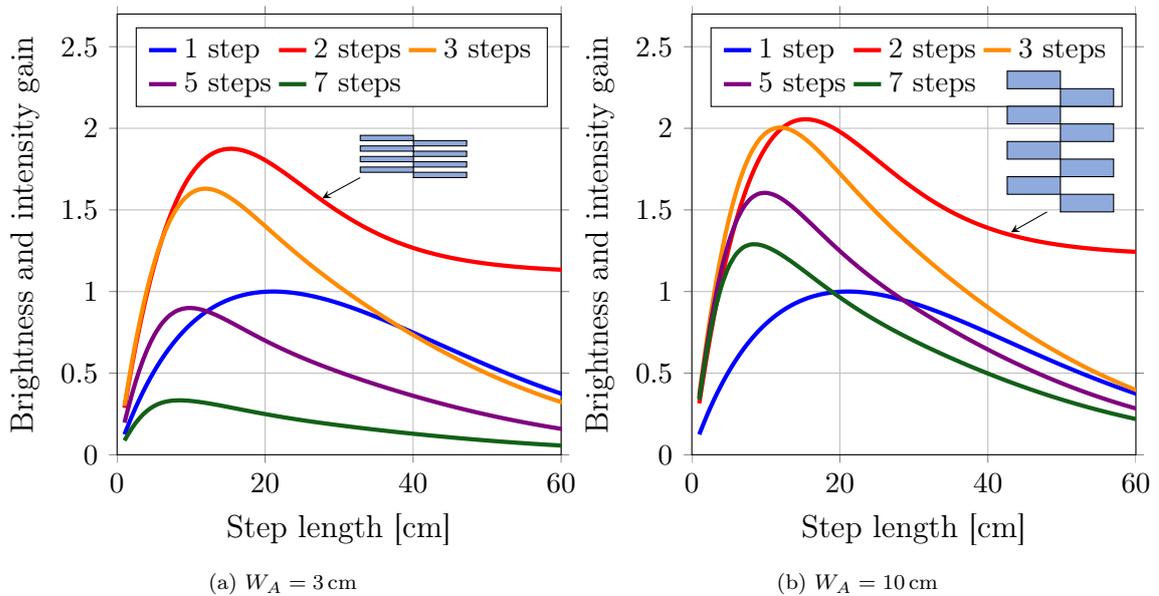

Note that the gains presented in this section may be further
influenced by the specifics of the moderator-reflector assembly and
should be evaluated through dedicated Monte Carlo simulations for a
particular neutron source, which will be addressed in forthcoming
publications.

Optimisation of the chessboard and staircase geometries for achieving
their maximal performance should include determination of the optimal
length and number of single moderators as well as specific gain
values.

\section{Conclusion and summary}
\label{sec:7}

Low-dimensional liquid para-hydrogen moderators, initially discovered and
extensively developed at the ESS, enhance the brightness of cold
neutrons compared to voluminous
moderators~\cite{Zanini2019a}. Although this enhancement comes at the
cost of reducing neutron beam intensity, it has been determined for
the ESS instrumental suite~\cite{Andersen2018} that the flat moderator
with the smallest dimension of \qty{3}{\cm}, providing a brightness
gain of about 2.5 with intensity losses of approximately
\qty{20}{\percent}, offers a reasonable trade-off when considering
various aspects of neutron beam extraction and transport to
instruments. In this work, we focused on high-aspect ratio rectangular
para-hydrogen moderators, where the width is much larger than its
height~(or vice versa). It was determined that due to the prevailing
single-collision mechanism of the thermal-to-cold neutron conversion
in para-hydrogen, the brightness gain for narrower moderators can reach
rather high values, approximately around~10~(see \figref{fig:20}).

However, these narrower moderators also exhibit even
greater intensity losses and are not ideal from the perspective of
neutron transport, mainly due to the insufficient illumination of the
entries to neutron guides~\cite{Konik2023a}.  To address this challenge, we have
introduced the concept of chessboard and staircase moderator
assemblies with extensively developed total surfaces.

These assemblies generate wide neutron beams of higher intensity while
preserving the enhanced brightness associated with high-aspect ratio
individual moderators. Though a cold neutron source built upon such
concept can principally provide an extremely high brightness, there is
a practical limit of achievable brightness gain imposed by various
factors including the moderator wall width and the limited volume of
the high-density thermal neutron region surrounding the reactor core
or spallation target. Analytic model calculations indicate that gains
of up to approximately~2.5 in both brightness and intensity are
achievable when compared to a source made of a single flat moderator
with the same width.  However, these gains will be affected by details
of the moderator-reflector assembly and should be estimated through
dedicated Monte Carlo simulations, which can only be conducted for a
particular neutron source and are beyond the scope of this general
study.

High brightness moderators assemblies can serve as not only primary
components in new neutron sources but also as upgraded alternatives
for the existing neutron guides in already established neutron
sources. This approach allows for the optimization of neutron beam
quality and intensity without needing to completely replace or rebuild
existing neutron guide systems. The brightness dependency on moderator
size is leading to the propagation of the ``one instrument~--- one
moderator'' concept, where ``brightness-hungry'' instruments are
placed on small moderators, while ``flux-hungry'' instruments are
placed on larger ones. Implementing this approach in practice is quite
demanding. The utilization of the suggested chessboard or staircase
configurations partially addresses these challenges, introducing a
novel ``universal'' moderator that is equally suitable for both
brightness- and flux-hungry instruments. However, the relatively large
length of moderator assemblies makes their application for short pulse
neutron sources very problematic.

We have also explored the concept of ``low dimensionality'' in
moderators, which yields a significant increase in brightness. We
demonstrated that achieving this requires moderators to be
low-dimensional both geometrically, implying a high aspect ratio, and
physically, indicating that the moderator's smallest dimension is
smaller than the triple mean free path of thermal neutrons in
para-hydrogen. This provides a physical explanation for the observed MCNP
result: when the high aspect ratio moderator is additionally
compressed along the longest direction, effectively giving it a
tube-like shape, the resulting increase in brightness is only
approximately \qty{20}{\percent}.

\section*{Acknowledgement}

This project was partially funded
from the European Union's
Horizon~2020 research and innovation programme under grant agreement
No.~871072.

\renewcommand\thesection{Appendix A}
\section{Total cold neutron intensity from rectangular moderator}
\setcounter{section}{0}
\setcounter{figure}{0}
\setcounter{table}{0}
\setcounter{equation}{0}
\renewcommand\thesection{A}
\renewcommand{\thefigure}{\thesection\arabic{figure}}
\renewcommand{\thetable}{\thesection\arabic{table}}
\renewcommand{\theequation}{\thesection\arabic{equation}}
\addcontentsline{toc}{section}{Appendix A}
\label{app:A}

\makeatletter
\@addtoreset{figure}{section}
\@addtoreset{table}{section}
\makeatother

Consider a moderator illuminated from the left
side~(\figref{fig:app:illumination:a}). The moderator area can be split into four
segments, each shaded in different colours. The neutron path length~$s_i$
in each segment depends on the position of the thermal neutron
emission point~$P$, which is offset from the center of the emitting surface
by a
distance~$\Delta$. The inclination angles~$\alpha_1$ and~$\alpha_2$ of
the lines connecting the neutron emission point~$P$ with the upper and
lower corners
\todo[author=KB]{angles???}
of the moderator are defined as follows:

\apptrue
\begin{figure}
\centering
\begin{subfigure}[B]{\figwidth}
\hspace*{0.2\linewidth}
\tikzsetnextfilename{main-figure72}
\begin{tikzpicture}
\end{tikzpicture}
\caption{Side wall illumination}
\ifapp
\label{fig:app:illumination:a}
\else
\label{fig:illumination:a}
\fi
\end{subfigure}
\begin{subfigure}[B]{\figwidth}
\hspace*{0.2\linewidth}
\tikzsetnextfilename{main-figure73}
\begin{tikzpicture}
\end{tikzpicture}
\caption{Bottom wall illumination}
\ifapp
\label{fig:app:illumination:b}
\else
\label{fig:illumination:b}
\fi
\end{subfigure}
\ifapp
\caption{Notations used for calculating partial cold neutron
intensities. Note that these diagrams do not depict angles
$\theta_3$ and $\theta_4$.}
\label{fig:app:illumination}
\else
\caption{Illustration of infinitely thin moderator illumination}
\label{fig:illumination}
\fi
\end{figure}

\begin{eqnarray}
\alpha_1(\Delta, W, H) & = & \arctan\left(\frac{H/2+\Delta}{W} \right)  \label{eq:a:1} \\
\alpha_2(\Delta, W, H) & = & \arctan\left(\frac{H/2-\Delta}{W} \right)  \label{eq:a:2}
\end{eqnarray}

\noindent The neutron path lengths $s_{i}$ in each segment are given as

\begin{eqnarray}
s_1 (\theta_1, W) & = & \dfrac{W}{\cos{\theta_1}}    \label{eq:a:3} \\
s_2 (\Delta, \theta_2, H) & = & \dfrac{H/2 + \Delta}{\sin{\theta_2}}    \label{eq:a:4} \\
s_3 (\theta_3, W) &  = & \dfrac{W}{\cos{\theta_3}}    \label{eq:a:5} \\
s_4 (\Delta, \theta_2, H) & = & \dfrac{H/2 - \Delta}{\sin{\theta_4}}    \label{eq:a:6}
\end{eqnarray}

\noindent where $\theta_{1-4}$ represent the angles between the normal
to the emitting surface and the corresponding neutron paths,
$s_{1-4}$.  Assuming a constant thermal neutron flux distribution over
the moderator surface, denoted as $B_{th0}$, the partial cold neutron
intensities from these four segments are given as follows:

\begin{eqnarray}
I_S^{(1)}(W,H) & = & B_{th0} \int_{-H/2}^{H/2} \int_0^{\alpha_1(\Delta, W, H)} \left(1 - \exp\Bigl(-\dfrac{s_1 (\theta_1, W)}{\Lambda_{th}}\Bigr) \right) \, d\theta \, d\Delta    \label{eq:a:7} \\
I_S^{(2)}(W,H) & = & B_{th0} \int_{-H/2}^{H/2} \int_{\alpha_1(\Delta, W, H)}^{\pi/2} \left(1 - \exp\Bigl(-\dfrac{s_2 (\Delta, \theta_2, H)}{\Lambda_{th}}\Bigr) \right) \, d\theta \, d\Delta   \label{eq:a:8} \\
I_S^{(3)}(W,H) & = & B_{th0} \int_{-H/2}^{H/2} \int_0^{\alpha_2(\Delta, W, H)} \left(1 - \exp\Bigl(-\dfrac{s_3 (\theta_3, W)}{\Lambda_{th}}\Bigr) \right) \, d\theta \, d\Delta    \label{eq:a:9}\\
I_S^{(4)}(W,H) & = & B_{th0} \int_{-H/2}^{H/2} \int_{\alpha_2(\Delta, W, H)}^{\pi/2} \left(1 - \exp\Bigl(-\dfrac{s_4 (\Delta, \theta_4, H)}{\Lambda_{th}}\Bigr) \right) \, d\theta \, d\Delta     \label{eq:a:10}
\end{eqnarray}

The total cold neutron intensity is equal to the sum of all partial intensities:

\begin{equation}
I_S(W,H)  =  I_S^{(1)}(W,H) + I_S^{(2)}(W,H) + I_S^{(3)}(W,H) + I_S^{(4)}(W,H) \label{eq:a:11}
\end{equation}

\bigskip

Consider the case where the moderator is illuminated from the
bottom~(\figref{fig:app:illumination:b}). It is important to note that
the only difference from the previously considered side illumination
is the interchange of the roles of height $H$ and width
$W$. Therefore, by substituting $W$ for $H$ in equations
(\ref{eq:a:1})--(\ref{eq:a:6}), one can immediately derive expressions
for the inclination angles $\alpha_1$ and $\alpha_2$:

\begin{eqnarray}
\beta_1(\Delta, W, H) & = & \arctan\left(\frac{W/2+\Delta}{H} \right)  \label{eq:a:12} \\
\beta_2(\Delta, W, H) & = & \arctan\left(\frac{W/2-\Delta}{H} \right)  \label{eq:a:13}
\end{eqnarray}

\noindent and for neutron path lengths $b_{i}$ in each segment:

\begin{eqnarray}
b_1 (\theta_1, H) & = & \dfrac{H}{\cos{\theta_1}}    \label{eq:a:14} \\
b_2 (\Delta, \theta_2, W) & = & \dfrac{W/2 + \Delta}{\sin{\theta_2}}    \label{eq:a:15} \\
b_3 (\theta_3, H) &  = & \dfrac{H}{\cos{\theta_3}}    \label{eq:a:16} \\
b_4 (\Delta, \theta_2, W) & = & \dfrac{W/2 - \Delta}{\sin{\theta_4}}    \label{eq:a:17}
\end{eqnarray}

\noindent where $\theta_{1-4}$ now represent the angles between the normal
to the emitting surface and the corresponding neutron paths,
$s_{1-4}$.

The partial cold neutron intensities from these four segments are given as follows:

\begin{eqnarray}
I_B^{(1)}(W,H) & = & B_{th0} \int_{-W/2}^{W/2} \int_0^{\beta_1(\Delta, W, H)} \left(1 - \exp\Bigl(-\dfrac{b_1 (\theta_1, H)}{\Lambda_{th}}\Bigr) \right) \, d\theta \, d\Delta    \label{eq:a:18} \\
I_B^{(2)}(W,H) & = & B_{th0} \int_{-W/2}^{W/2} \int_{\beta_1(\Delta, W, H)}^{\pi/2} \left(1 - \exp\Bigl(-\dfrac{b_2 (\Delta, \theta_2, W)}{\Lambda_{th}}\Bigr) \right) \, d\theta \, d\Delta   \label{eq:a:19} \\
I_B^{(3)}(W,H) & = & B_{th0} \int_{-W/2}^{W/2} \int_0^{\beta_2(\Delta, W, H)} \left(1 - \exp\Bigl(-\dfrac{b_3 (\theta_3, H)}{\Lambda_{th}}\Bigr) \right) \, d\theta \, d\Delta    \label{eq:a:20}\\
I_B^{(4)}(W,H) & = & B_{th0} \int_{-W/2}^{W/2} \int_{\beta_2(\Delta, W, H)}^{\pi/2} \left(1 - \exp\Bigl(-\dfrac{b_4 (\Delta, \theta_4, W)}{\Lambda_{th}}\Bigr) \right) \, d\theta \, d\Delta     \label{eq:a:21}
\end{eqnarray}

The total cold neutron intensity for the bottom illumination is equal to the sum of these partial intensities:

\begin{equation}
I_B(W,H)  =  I_B^{(1)}(W,H) + I_B^{(2)}(W,H) + I_B^{(3)}(W,H) + I_B^{(4)}(W,H). \label{eq:a:22}
\end{equation}

\renewcommand\thesection{Appendix B}
\section{Shadowing effect}
\setcounter{section}{0}
\setcounter{figure}{0}
\setcounter{table}{0}
\setcounter{equation}{0}
\renewcommand\thesection{B}
\renewcommand{\thefigure}{\thesection\arabic{figure}}
\renewcommand{\thetable}{\thesection\arabic{table}}
\renewcommand{\theequation}{\thesection\arabic{equation}}
\addcontentsline{toc}{section}{Appendix B}
\label{app:B}

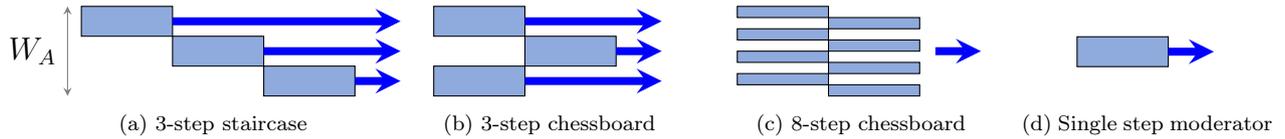
\begin{figure}
\centering
\def\width{12mm}
\def\height{0.33*\width}
\begin{subfigure}{0.28\linewidth}
\hspace*{-7mm}
\centering
\tikzsetnextfilename{main-figure74}
\begin{tikzpicture}
\end{tikzpicture}
\caption{3-step staircase}
\label{fig:shadow:layouts:staircase}
\end{subfigure}
\begin{subfigure}{0.24\linewidth}
\centering
\tikzsetnextfilename{main-figure75}
\begin{tikzpicture}
\end{tikzpicture}
\caption{3-step chessboard}
\label{fig:shadow:layouts:chessboard}
\end{subfigure}
\begin{subfigure}{0.24\linewidth}
\centering
\tikzsetnextfilename{main-figure76}
\begin{tikzpicture}
\end{tikzpicture}
\caption{8-step chessboard}
\label{fig:shadow:layouts:4}
\end{subfigure}
\begin{subfigure}{0.2\linewidth}
\centering
\tikzsetnextfilename{main-figure77}
\begin{tikzpicture}
\end{tikzpicture}
\caption{Single step moderator}
\label{fig:shadow:layouts:single}
\end{subfigure}
\caption{Layouts of moderator assemblies used to estimate the shadowing effect. Blue arrows indicate outgoing cold neutron beam direction.}
\label{fig:shadow:layouts}
\end{figure}

To estimate the shadowing effect of neighbouring steps, we compare the
brightness of a single para-hydrogen step positioned alone in the centre of a
long neutron beam extraction channel with that of the staircase or
chessboard configuration~(\figref{fig:shadow:layouts}), where each
individual step has a size of~($W_A/3 \times 10 \times
10$)~\unit{\cm\cubed}.

\begin{figure}
\def\thin{3}
\foreach \WA in {3,10} {\begin{subfigure}{\figwidth}
\tikzsetnextfilename{main-figure78}
\begin{tikzpicture}
\end{tikzpicture}
\caption{\ifx\WA\thin $W_A = \qty{3}{\cm}$ \else $W_A = \qty{10}{\cm}$ \fi}
\label{fig:shadowing:ratio:\WA}
\end{subfigure}
}
\caption{Shadowing factors as functions of neutron wavelength for selected configurations.}
\label{fig:shadowing:ratio}
\end{figure}
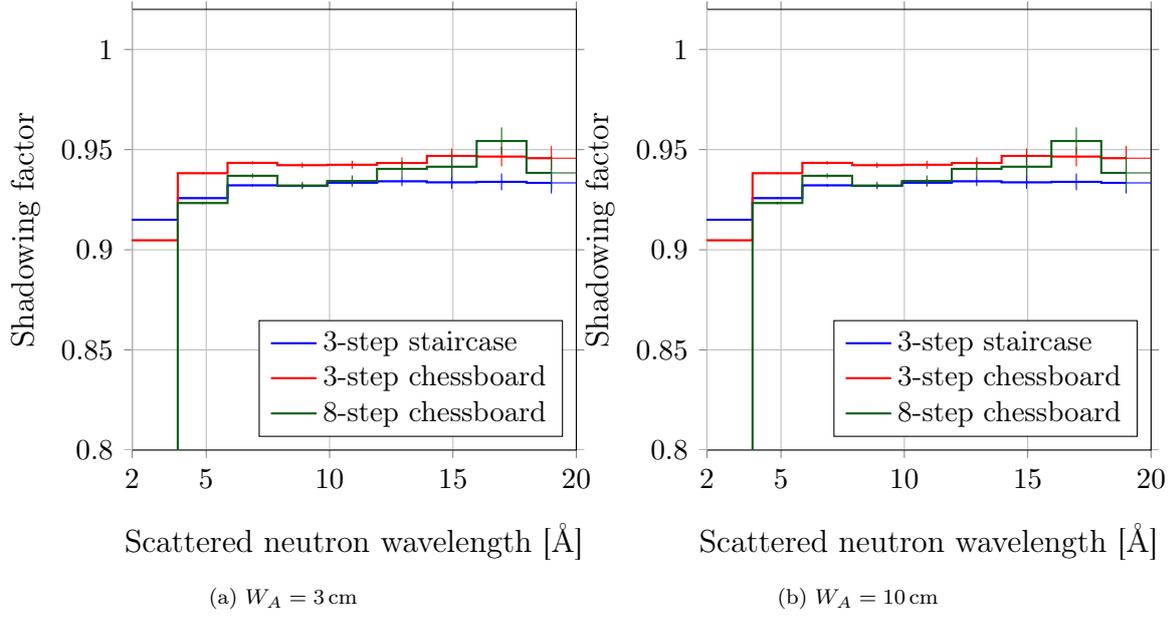

The brightness spectra were calculated and utilised to determine
integrated brightness values across various wavelength
ranges~(\tabref{tab:shadow_results1}).  Since the MFP for neutrons
with wavelength \qtyrange{4}{20}{\angstrom} is larger than for
neutrons with wavelength \qtyrange{3}{4}{\angstrom}~(see
\figref{fig:8:b}), the effect of shadowing for the former is smaller.

The shadowing factors are then calculated as the loss of brightness
with respect to the single step moderator. Examples are shown in
Fig.~\ref{fig:shadowing:ratio} and \tabref{tab:shadow_results2} for
various wavelength ranges. As can be seen, the shadowing for cold
neutrons with wavelengths in the range of \qtyrange{4}{20}{\angstrom}
is less than \qty{8}{\percent} for both discussed moderator
widths. Since we did not conduct the MCNP calculations for all
possible staircase and chessboard configurations discussed
in~\secref{sec:6}, we will assume a conservative estimate for the
shadowing factor as $F_{shadow} = \qty{10}{\percent}$ and use it for
calculations. Note, that in the range of shorter wavelengths for the
narrow, \qty{3}{\cm} moderator assembly, the shadowing is slightly higher,
\qty{15}{\percent}.

\begin{table}
\centering
\begin{tabular}{cc|cc|cc}
\toprule
& & \multicolumn{2}{c|}{$W_A=\qty{3}{\cm}$} & \multicolumn{2}{c}{$W_A=\qty{10}{\cm}$} \\
\midrule
Layout & Figure & \qtyrange{3}{4}{\angstrom} & \qtyrange{4}{20}{\angstrom} & \qtyrange{3}{4}{\angstrom} & \qtyrange{4}{20}{\angstrom} \\
\midrule
3-step staircase & \ref{fig:shadow:layouts:staircase}  & \num{1.452 +- 0.003} & \num{1.842 +- 0.003} & \num{1.242 +- 0.002} & \num{1.652 +- 0.002}\\
3-step chessboard & \ref{fig:shadow:layouts:chessboard} & \num{1.462 +- 0.003} & \num{1.865 +- 0.003} & \num{1.247 +- 0.002} & \num{1.646 +- 0.002}\\
8-step chessboard & \ref{fig:shadow:layouts:4}  & \num{1.349 +- 0.003} & \num{1.837 +- 0.003} & \num{1.212 +- 0.002} & \num{1.690 +- 0.002}\\
Single step & \ref{fig:shadow:layouts:single} & \num{1.578 +- 0.003} & \num{1.984 +- 0.003} & \num{1.302 +- 0.002} & \num{1.701 +- 0.002}\\
\bottomrule
\end{tabular}
\caption{Central step brightness [\unit{\per\angstrom\per\cm\squared\per\steradian}] for different moderator layouts.}
\label{tab:shadow_results1}
\todo[color=yellow]{Update statistics}
\end{table}

\begin{table}
\centering
\begin{tabular}{cc|cc|cc}
\toprule
& & \multicolumn{2}{c|}{$W_A=\qty{3}{\cm}$} & \multicolumn{2}{c}{$W_A=\qty{10}{\cm}$} \\
\midrule
Layout & Figure & \qtyrange{3}{4}{\angstrom} & \qtyrange{4}{20}{\angstrom} & \qtyrange{3}{4}{\angstrom} & \qtyrange{4}{20}{\angstrom} \\
\midrule
3-step staircase & \ref{fig:shadow:layouts:staircase} & \qty{8.23 +- 0.04}{\percent} & \qty{7.17 +- 0.03}{\percent} & \qty{4.77 +- 0.03}{\percent} & \qty{2.89 +- 0.03}{\percent} \\
3-step chessboard & \ref{fig:shadow:layouts:chessboard}  & \qty{7.39 +- 0.04}{\percent} & \qty{5.99 +- 0.04}{\percent} & \qty{4.48 +- 0.03}{\percent} & \qty{3.28 +- 0.03}{\percent} \\
8-step chessboard & \ref{fig:shadow:layouts:4} & \qty{14.27 +- 0.06}{\percent} & \qty{7.14 +- 0.05}{\percent} & \qty{6.99 +- 0.05}{\percent} & \qty{0.55 +- 0.04}{\percent} \\
\bottomrule
\end{tabular}
\caption{Shadowing factors for staircase and chessboard arrangements.}
\label{tab:shadow_results2}
\end{table}

\newpage
\bibliographystyle{elsarticle-num}
\bibliography{references}

\begin{thebibliography}{10}
\expandafter\ifx\csname url\endcsname\relax
  \def\url#1{\texttt{#1}}\fi
\expandafter\ifx\csname urlprefix\endcsname\relax\def\urlprefix{URL }\fi
\expandafter\ifx\csname href\endcsname\relax
  \def\href#1#2{#2} \def\path#1{#1}\fi

\bibitem{kiyanagi2003}
Y.~Kiyanagi, M.~Ooi, H.~Ogawa, M.~Furusaka, Development of hydrogen cold
  moderator systems for a spallation neutron source, Journal of Neutron
  Research 11~(1--2) (2003) 3--11.
\newblock \href {https://doi.org/10.1080/1023816031000100860}
  {\path{doi:10.1080/1023816031000100860}}.

\bibitem{ageron1969}
P.~Ageron, P.~De~Beaucourt, H.~Harig, A.~Lacaze, M.~Livolant, Experimental and
  theoretical study of cold neutron sources of liquid hydrogen and liquid
  deuterium, Cryogenics 9~(1) (1969) 42--50.
\newblock \href {https://doi.org/10.1016/0011-2275(69)90257-4}
  {\path{doi:10.1016/0011-2275(69)90257-4}}.

\bibitem{shin2010}
Y.~Shin, C.~Lavelle, W.~M. Snow, D.~V. Baxter, X.~Tong, H.~Yan, M.~Leuschner,
  Measurements of the neutron brightness from a phase {II} solid methane
  moderator at the {LENS} neutron source, Nuclear Instruments and Methods in
  Physics Research Section A: Accelerators, Spectrometers, Detectors and
  Associated Equipment 620~(2--3) (2010) 375--381.
\newblock \href {https://doi.org/10.1016/j.nima.2010.03.108}
  {\path{doi:10.1016/j.nima.2010.03.108}}.

\bibitem{Batkov2013a}
K.~Batkov, A.~Takibayev, L.~Zanini, F.~Mezei, Unperturbed moderator brightness
  in pulsed neutron sources, Nuclear Instruments and Methods in Physics
  Research Section A: Accelerators, Spectrometers, Detectors and Associated
  Equipment 729 (2013) 500--505.
\newblock \href {https://doi.org/10.1016/j.nima.2013.07.031}
  {\path{doi:10.1016/j.nima.2013.07.031}}.

\bibitem{Mezei2013a}
F.~Mezei, L.~Zanini, A.~Takibayev, K.~Batkov, E.~Klinkby, E.~Pitcher,
  T.~Sch\"onfeldt, Low dimensional neutron moderators for enhanced source
  brightness, Journal of Neutron Research 17~(2) (2014) 101--105.
\newblock \href {https://doi.org/10.3233/JNR-140013}
  {\path{doi:10.3233/JNR-140013}}.

\bibitem{Zanini2019a}
L.~Zanini, K.~Andersen, K.~Batkov, F.~Mezei, A.~Takibayev, E.~Klinkby,
  T.~Sch\"onfeldt, Design of the cold and thermal neutron moderators for the
  {E}uropean {S}pallation {S}ource, Nuclear Instruments and Methods in Physics
  Research Section A: Accelerators, Spectrometers, Detectors and Associated
  Equipment 925 (2019) 33--52.
\newblock \href {https://doi.org/10.1016/j.nima.2019.01.003}
  {\path{doi:10.1016/j.nima.2019.01.003}}.

\bibitem{Batkov2013b}
K.~Batkov, A.~Takibayev, L.~Zanini, F.~Mezei, Cold moderators for long pulse
  neutron sources: unperturbed brightness, in: International Topical Meeting on
  Nuclear Applications of Accelerators (AccApp--XI), Bruges, Belgium, 2013.

\bibitem{SNSSTS}
F.~X. Gallmeier, I.~Remec, A liquid hydrogen tube moderator arrangement for
  {SNS} second target station, Review of Scientific Instruments 93~(8) (2022).
\newblock \href {https://doi.org/10.1063/5.0095900}
  {\path{doi:10.1063/5.0095900}}.

\bibitem{HBS}
U.~R{\"u}cker, T.~Cronert, J.~Voigt, J.~Dabruck, P.-E. Doege, J.~Ulrich,
  R.~Nabbi, Y.~Be{\ss}ler, M.~Butzek, M.~B{\"u}scher, et~al., The {J}{\"u}lich
  high-brilliance neutron source project, The European Physical Journal Plus
  131~(1) (2016) 19.
\newblock \href {https://doi.org/10.1140/epjp/i2016-16019-5}
  {\path{doi:10.1140/epjp/i2016-16019-5}}.

\bibitem{moskvin2023}
E.~Moskvin, N.~Grigoryeva, N.~Kovalenko, S.~Grigoryev, Conceptual design of a
  time-of-flight powder diffractometer for a compact neutron source, Journal of
  Surface Investigation: X-ray, Synchrotron and Neutron Techniques 17~(4)
  (2023) 804--809.
\newblock \href {https://doi.org/10.1134/S1027451023040109}
  {\path{doi:10.1134/S1027451023040109}}.

\bibitem{Konik2023a}
P.~Konik, A.~Ioffe, A new method to find out the optimal neutron moderator size
  based on neutron scattering instrument parameters, Nuclear Instruments and
  Methods in Physics Research Section A: Accelerators, Spectrometers, Detectors
  and Associated Equipment 1056 (2023) 168643.
\newblock \href {https://doi.org/10.1016/j.nima.2023.168643}
  {\path{doi:10.1016/j.nima.2023.168643}}.

\bibitem{Andersen2018}
K.~H. Andersen, M.~Bertelsen, L.~Zanini, E.~B. Klinkby, T.~Sch{\"o}nfeldt,
  P.~M. Bentley, J.~Saroun, Optimization of moderators and beam extraction at
  the {ESS}, Journal of applied crystallography 51~(2) (2018) 264--281.
\newblock \href {https://doi.org/10.1107/S1600576718002406}
  {\path{doi:10.1107/S1600576718002406}}.

\bibitem{goorley2012}
J.~T. Goorley, et~al., Initial {MCNP6} {R}elease {O}verview, Nuclear
  Technology~(180) (2012) 298--315.
\newblock \href {https://doi.org/10.13182/NT11-135}
  {\path{doi:10.13182/NT11-135}}.

\bibitem{mattauch2017}
S.~Mattauch, A.~Ioffe, D.~Lott, L.~Bottyán, J.~Daillant, M.~Markó,
  A.~Menelle, S.~Sajti, T.~Veres, {HERITAGE}: the concept of a giant flux
  neutron reflectometer for the exploration of 3-d structure of free-liquid and
  solid interfaces in thin films, Nuclear Instruments and Methods in Physics
  Research Section A: Accelerators, Spectrometers, Detectors and Associated
  Equipment 841 (2017) 34--46.
\newblock \href {https://doi.org/10.1016/j.nima.2016.09.043}
  {\path{doi:10.1016/j.nima.2016.09.043}}.

\bibitem{Ikeda2009}
Y.~Ikeda, {J-PARC} status update, Nuclear Instruments and Methods in Physics
  Research Section A: Accelerators, Spectrometers, Detectors and Associated
  Equipment 600~(1) (2009) 1--4.
\newblock \href {https://doi.org/10.1016/j.nima.2008.11.019}
  {\path{doi:10.1016/j.nima.2008.11.019}}.

\bibitem{Watanabe2003}
N.~Watanabe, Neutronics of pulsed spallation neutron sources, Reports on
  Progress in Physics 66~(3) (2003) 339.
\newblock \href {https://doi.org/10.1088/0034-4885/66/3/202}
  {\path{doi:10.1088/0034-4885/66/3/202}}.

\bibitem{conlin2012}
J.~L. Conlin, D.~K. Parsons, F.~B. Brown, R.~E. MacFarlane, R.~C. Little, M.~C.
  White, Continuous ${S}(\alpha,\beta)$ {C}apability in {MCNP}, in: ANS Annual
  Meeting, Chicago, IL, USA, 2012.

\bibitem{chadwick2006}
M.~Chadwick, P.~Oblo\v{z}insk\`y, M.~Herman, N.~Greene, R.~McKnight, D.~Smith,
  P.~Young, R.~MacFarlane, G.~Hale, S.~Frankle, A.~Kahler, T.~Kawano,
  R.~Little, D.~Madland, P.~Moller, R.~Mosteller, P.~Page, P.~Talou,
  H.~Trellue, M.~White, W.~Wilson, R.~Arcilla, C.~Dunford, S.~Mughabghab,
  B.~Pritychenko, D.~Rochman, A.~Sonzogni, C.~Lubitz, T.~Trumbull, J.~Weinman,
  D.~Brown, D.~Cullen, D.~Heinrichs, D.~McNabb, H.~Derrien, M.~Dunn, N.~Larson,
  L.~Leal, A.~Carlson, R.~Block, J.~Briggs, E.~Cheng, H.~Huria, M.~Zerkle,
  K.~Kozier, A.~Courcelle, V.~Pronyaev, S.~van~der Marck, {ENDF/B-VII.0}:
  {N}ext generation evaluated nuclear data library for nuclear science and
  technology, Nuclear Data Sheets 107~(12) (2006) 2931--3059.
\newblock \href {https://doi.org/10.1016/j.nds.2006.11.001}
  {\path{doi:10.1016/j.nds.2006.11.001}}.

\bibitem{Wurz1973}
H.~W\"urz, {U}ntersuchungen zur {N}eutronenthermalisierung an fl\"ussigem
  {O}rtho- und {P}ara-{W}asserstoff, 1973, {KFK} 1697,
  \href{https://d-nb.info/1187253243/34}{https:/\!/d-nb.info/1187253243/34}.

\end{thebibliography}

\end{document}